\documentclass[aoas,nameyear,MSNbibl,dvips]{arximspdf}
\usepackage{graphicx}

\doi{10.1214/10-AOAS452}
\volume{5}
\issue{2B}
\pubyear{2011}
\firstpage{1328}
\lastpage{1359}

\makeatletter
\newcommand{\bmu}{\bolds\mu}
\renewcommand{\epsilon}{\varepsilon}
\newcommand{\bepsilon}{\bolds\epsilon}
\newcommand{\bY}{\mathbf{Y}}
\newcommand{\bT}{\mathbf{T}}
\makeatother

\begin{document}
\begin{frontmatter}

\title{Bayesian hierarchical modeling for temperature reconstruction
from geothermal data\thanksref{T1}}
\runtitle{Temperature reconstruction}

\begin{aug}
\author[A]{\fnms{Jenn\'{y}} \snm{Brynjarsd\'{o}ttir}\corref{}\ead[label=e1]{brynjarsdottir.1@osu.edu}}
\and
\author[A]{\fnms{L. Mark} \snm{Berliner}\ead[label=e2]{mb@stat.osu.edu}}
\runauthor{J. Brynjarsd\'{o}ttir and L. M. Berliner}
\pdfauthor{Jenny Brynjarsdottir, L. Mark Berliner}
\affiliation{Ohio State University}
\address[A]{Department of Statistics \\
Ohio State University \\ 1958 Neil Avenue \\ 404 Cockins Hall\\
Columbus, Ohio 43210 \\ USA \\ \printead{e1}\\
\hphantom{\textsc{E-mail:} }\printead*{e2}} 
\end{aug}
\thankstext{T1}{Supported by NSF Grant
ATM-07-24403.}

\received{\smonth{6} \syear{2010}}
\revised{\smonth{12} \syear{2010}}

%
\begin{abstract}
We present a Bayesian hierarchical modeling approach to paleoclimate
reconstruction using borehole temperature profiles. The
approach relies on modeling heat conduction in solids via the heat
equation with step function, surface boundary conditions. Our analysis
includes model error and assumes that the boundary conditions are
random processes. The formulation also enables separation of
measurement error and model error.
We apply the analysis to data from nine borehole temperature
records from the San Rafael region in Utah.
We produce ground surface temperature histories with uncertainty
estimates for the
past 400 years. We pay special attention to use of prior parameter models
that illustrate borrowing strength in a combined analysis for all nine
boreholes. In addition, we review selected sensitivity analyses.
\end{abstract}

%

\end{frontmatter}

\section{Introduction}\label{sec1}
Reconstruction of past climate plays an important role in
climate change analysis. Comparisons between
climate behavior before and after human influences are a relevant
component of claims of attribution of
climate change to our activities. The term \textit{paleoclimate} is used
in reference to data analysis and modeling of climate for times before
the modern era of data collection. The time periods of
interest range from hundreds to millions of years before the present.
Of course, as our
interest moves toward the past, the availability of reliable and
spatially and temporally plentiful observations of weather and climate
diminishes. In response, scientists have developed the use of \textit
{proxy} indicators of climate [\citet{JansenEtal2007}].

A proxy is a quantity
taking values that respond to climate behavior. For example, annual
tree ring thicknesses respond to weather
variables that control growth, that is, temperature and
precipitation. If one develops useful models for proxies as rough
functions of climate, then the models can be inverted to estimate
climate behavior based on observed proxy variables. The statistical
notions of regression and inverse regression analyses are immediately
evident.
Though exceptions exist and the trend is positive
[e.g., \citet{HasletEtal2006}; \citeauthor{LiNychkaAmmann2007}
(\citeyear{LiNychkaAmmann2007,LiNychkaAmmann2010})],
there has been insufficient participation by statisticians in
paleoclimate reconstruction. This is surprising in view of the
richness of the statistical challenges: proxies are themselves
observed with error; the forward models for proxies as functions of
climate are partially known at best and subject to model errors;
inverse analyses are not trivial statistically; spatial and temporal
coverage and mismatches between proxy data sets and desired climate
inferences are among the issues. Furthermore, there is substantial
interest in paleoclimate among policy makers
and the general public
[\citet{WegmanScottSaid2006}; \citet{SmithBerlinerGuttorp2010}].

In this article we focus on the critical problem of
surface temperature reconstruction [\citet{JansenEtal2007}; \citet{NorthEtal2006}].
We analyze \textit{borehole temperature} data sets and their use in
reconstructing surface temperature time series
[e.g., \citet{BeltramiMareschal1995}; \citet{PollackHuangShen1998}].
A borehole is a narrow shaft drilled into
the ground (or ice), typically vertically, in search of
subterranean resources (gas, oil, water, minerals, etc.). Boreholes
are also used to monitor environmental processes (e.g.,
percolation of contaminants) or as pilots to access suitability for
more intense drilling or construction projects.
Borehole data that are used for temperature reconstruction are
typically obtained as byproducts of such projects.
Therefore, borehole data are
observations of opportunity rather than
having been designed with climate reconstruction in mind.
We note that borehole data are temperature measurements, and, hence,
perhaps not as indirect a~measure of surface temperature as other proxies.
However, the problems of developing and
inverting a model for borehole data as a function of surface
temperatures are challenging.

The underlying theory for using borehole data to infer surface
temperature is the physics of heat conduction. In principle, the
transfer of heat is governed by the \textit{heat equation}. This
is a partial differential equation describing the temporal evolution
of the temperature field over some domain. The idea is that the
surface temperatures over time serve as boundary conditions for the
evolution of temperature below the surface. Then information
regarding subsurface temperatures can be inverted to estimate the
boundary conditions.\looseness=-1

There are important issues and
uncertainties that arise in applying this strategy. First, subsurface
temperatures respond to ground-surface temperatures as opposed to near
surface air temperatures. Though the latter two are related, they are
not identical, perhaps due to snow cover and other factors.
Next, as heat conducts
into deepening levels of the subsurface, it spreads or smears out,
leading to losses in information regarding the boundary
as time increases. Though this problem is well known,
we will seek explicit characterizations of this loss of
information as reflected in
uncertainty measures associated with our results. Another
issue is that there are factors affecting heat conduction that are
difficult to quantify. For example, conduction rates depend on
characteristics of the media (i.e., rock types) through which the heat
flows. Further, percolation of water through the media also impacts
heat flow. For such reasons, we incorporate the heat equation with
error and unknown parameters in our modeling.

We present Bayesian modeling and analysis for data from 9 boreholes
in the San Rafael region in Utah.
To combine information from
these boreholes, we assume model parameters are site-specific, but
sampled from common distributions.
Our modeling incorporates both the observations and
physics into an analysis that is sensitive to the uncertainties in
both information sources. The use of such \textit{physical-statistical}
analyses in geophysical problems is increasing; for examples, see
\citet{Berliner2003}, \citet{BerlinerEtal2008} and
\citet{WikleMilliffEtal2001}. See \citet{HopcroftGallagherPain2007} for a
related Bayesian analysis of borehole data.\vspace*{-4pt}

\subsection{Review: Borehole data analysis}\label{sec1.1}
The conventional approach is to frame analyses
in terms of
\textit{reduced} temperatures defined as follows.
For a~given borehole, consider $N$
depths $z_1,\ldots ,z_N$,
where increasing values of $z$ correspond to increasing depths. Let
$\mathbf{T}$ be the $N \times1$ vector of true temperatures at these
depths. The corresponding vector of reduced temperatures is given
by\vspace*{-4pt}
%
\begin{equation}
\label{eq:Reduced}
\mathbf{T}_r = \mathbf{T} - T_0 \mathbf{1} - q_0 \mathbf{R},
\vspace*{-2pt}
\end{equation}
where $T_0$
is the \textit{surface temperature intercept}, $\mathbf{1}$ is an
$N \times1$ vector
whose elements are all equal to one, $q_0$ represents
\textit{background heat flow}, and $\mathbf{R}$ is an $N \times1$
vector of
\textit{thermal resistances} at each of the depths. This modeling step is
intended to account for the fact that both heating from the earth's
core and rates of heat conduction vary with depth, thereby justifying
use of the simple heat equation model described in
Section~\ref{subsec:physics}.

The thermal resistances account for differences in heat conduction and are
assumed known throughout the analysis (see Section \ref{sec:data}).
To deal with the unknowns $T_0$
and $q_0$, it is customary to replace the true temperatures $\mathbf{T}$
by the observed temperatures, $\mathbf{Y}$, in
(\ref{eq:Reduced}). Then, $T_0$ and $q_0$ are estimated
via least squares by regressing $\mathbf{Y}$ onto $\mathbf{R}$.
In that step a subset of the data is used, corresponding
to those depths where the climate change signal is assumed to be negligible
[for our data this means below 150~m or 200~m, depending on
the region; \citet{HarrisChapman1995}]. The resulting estimates\vspace*{-2pt}
%
\begin{equation} \label{eq:Reducedcoor}
\hat{T}_r(z_i) = Y(z_i) - \bigl ( \hat{T}_0 + \hat{q}_0
R(z_i) \bigr),\qquad i = 1,\ldots ,N,
\end{equation}
are then treated as the true reduced temperatures.

Having made the above adjustments, the heat equation is assumed to
apply. Let $\mathbf{T}_h$ be a $K \times1$ surface
temperature history vector. Here, the surface temperatures are
assumed to be constants over $K$ time intervals used in the analysis.
The heat equation can be solved (see Section \ref{sec:Methods}),
leading to the linear relationship
%
\begin{equation}
\label{eq:model1}
\hat{\mathbf{T}}_r = A \mathbf{T}_h,
\end{equation}
where $A$ is an $N \times K$ matrix developed from the solution to the
heat equation [see (\ref{eq:vectorsol})]. The objective then is to solve
the inverse problem, that is, obtain an estimate of $\bT_h$.
In most examples, $A$
is ill-conditioned and the inversion of $A^\prime A$ is unstable.
Therefore, traditional
regression methods which involve taking the inverse of $A^\prime A$
lead to unstable estimates of $\bT_h$. A common approach is to use a
singular value decomposition (SVD) of the $A$ matrix and retain only a
few of the singular
vectors [\citet{VasseurEtal1983}; \citet{BeltramiMareschal1991}; \citet{MareschalBeltrami1992}; \citet{HarrisChapman1995}].
In this paper we take a hierarchical Bayesian regression
approach that does not involve taking the inverse of $A^\prime A$.
Hence, we avoid
having to pick the number of singular
vectors (principal components) to retain.

Other approaches can be found in the geophysical literature.
For example, functional space inversion is a popular method
[\citeauthor{ShenBeck1991} (\citeyear{ShenBeck1991,ShenBeck1992}); \citet{HarrisChapman1998a}]. A comparative
study of
some inverse methods in this setting can be found in \citet{ShenEtal1992}.

\subsection{Outline}
The paper is organized as follows. In Section \ref{sec:data} we
describe the
borehole data used in the analysis. In Section \ref{sec:Methods}
after a brief introduction to the physical model the analysis is
based on, we develop a Bayesian
hierarchical model. To best convey the ideas, we first
present a single-site borehole model in Section~\ref{sec:SingleSite}.
We extend it to include data from multiple boreholes in
Section \ref{sec:MultSite}.
The results are presented in Section \ref{sec:Results}.
In Section~\ref{sec:ResultsSingle} we
compare the results to those of single-site models and
in Section \ref{sec:Sens} we present
a number of sensitivity analysis. We end with a discussion in
Section \ref{sec:Discussion}.

\begin{figure}

\includegraphics{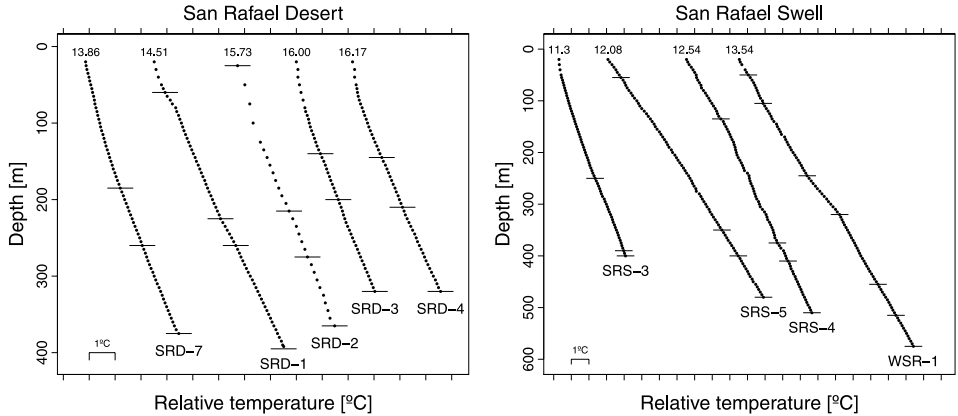}

\caption{Measured temperature-depth profiles from
boreholes in the San Rafael Desert \textup{(left)} and
the San Rafael Swell \textup{(right)}. The temperatures are shifted so they do
not overlap, one tick on the x axis corresponds to $1^{\circ}$C.
The value of the shallowest measurement is shown above each profile.
The horizontal line segments show the formation boundaries, the names
of the formations are given in Table~\protect\ref{tab:Cond}.}\label{fig:Data}
\end{figure}

\section{Data} \label{sec:data}\label{sec2}
We consider borehole data from the Colorado Plateau in Utah.
The data consist of nine measured temperature-depth profiles
(shown in Figure \ref{fig:Data})
belonging to two regions, the San Rafael Desert and the San Rafael Swell.
The geography of these regions is characterized by layered sedimentary
rocks that each have different thermal conductivities.
Measurements of these different conductivities are available
[\citet{BodellChapman1982}]
and have been adjusted to the specific formations in the boreholes
used in this analysis so that the
estimated thermal conductivity for each formation may be different
between regions but not
within regions [see \citet{HarrisChapman1995}
for details about these adjustments].
The adjusted thermal conductivities~($k$) for each sedimentary
formation and abbreviated formation names are shown
in Table \ref{tab:Cond} along with the formation boundaries
within each borehole.
For definitions of the
abbreviated formation names see \citet{BodellChapman1982}.
The sedimentary formation boundaries are also
shown as small horizontal line segments in
Figure \ref{fig:Data}.
For more background on the data and temperature reconstructions based
on them
see \citeauthor{HarrisChapman1995} (\citeyear{HarrisChapman1995,HarrisChapman1998a}).

Thermal resistance ($R$) is a function of depth and the thermal
conductivity at that depth, and
is calculated as
%
\begin{equation} \label{eq:ThermRes}
R(z_i) = \sum_{l=1}^i \frac{z_l - z_{l-1}}{k(z_l)}, \qquad i = 1, \ldots
, N,
\end{equation}
where $k(z_l)$ is the conductivity for the depth interval from
$z_{l-1}$ to $z_l$.
Recall that $z_0=0$ denotes the surface and $z$ is increasing in depth.
If the conductivity is constant ($k$) for the entire borehole,
(\ref{eq:ThermRes}) reduces to $R(z_i) = z_i/k$, that is,
a constant temperature gradient.
%
\begin{table}
\tabcolsep=0pt
\tablewidth=300pt
\caption{Depth to formation boundaries within each
borehole, names of
the formation types down to each boundary and the corresponding thermal
conductivities ($k$).
The table also shows for each borehole the estimated surface
temperature intercept,~$\hat{T}_0$,~and the year the borehole was logged}\label{tab:Cond}
\begin{tabular*}{300pt}{@{\extracolsep{\fill}}lccc@{}}
\hline
\textbf{Borehole} & \textbf{Depth [m]} & \textbf{Formation} &$\bolds k$ \textbf{[W/mK]}  \\
\hline
\multicolumn{4}{@{}l@{}}{\textit{San Rafael Desert}}\\
 \quad \textit{SRD-1}      & 60   & Jca  & 2.91\\
 \quad $\hat{T}_0 = 13.72$ & 225  & Jna  & 4.09\\
 \quad year: 1979          & 260  & JTrk & 3.96\\
                    & 395  & Trwi & 3.86\\[5pt]
 \quad \textit{SRD-2}      & 25   & Jca  & 2.91\\
 \quad $\hat{T}_0 = 15.12$ & 215  & Jna  & 4.09\\
 \quad year: 1976          & 275  & JTrk & 3.96\\
                    & 365  & Trwi & 3.86\\[5pt]
 \quad \textit{SRD-3}      & 140  & Jna  & 4.09\\
 \quad $\hat{T}_0 = 15.38$ & 200  & JTrk & 3.96\\
 \quad year: 1979          & 320  & Trwi & 3.86\\[5pt]
 \quad \textit{SRD-4}      & 145  & Jna  & 4.09\\
 \quad $\hat{T}_0 = 15.51$ & 210  & JTrk & 3.96\\
 \quad year: 1979          & 320  & Trwi & 3.86\\[5pt]
 \quad \textit{SRD-7}      & 185  & Jna  & 4.09\\
 \quad $\hat{T}_0 = 13.16$ & 260  & JTrk & 3.96\\
 \quad year: 1980          & 375  & Trwi & 3.86\\[3pt]
\multicolumn{4}{@{}l@{}}{\textit{San Rafael Swell}} \\
 \quad \textit{SRS-3}      & 250 & Pco  & 5.01 \\
 \quad $\hat{T}_0 = 10.76$ & 390 & Pec  & 4.35 \\
 \quad year: 1979          & 400 & Mr   & 4.82 \\[5pt]
 \quad \textit{SRS-4}      & 135 & Jca  & 2.91 \\
 \quad $\hat{T}_0 =11.82$  & 375 & Jna  & 4.18 \\
 \quad year: 1979          & 410 & JTrk & 3.86 \\
                    & 510 & Trwi & 4.17 \\[5pt]
 \quad \textit{SRS-5}      & 55  & Jca  & 2.91 \\
 \quad $\hat{T}_0 =11.82$  & 350 & Jna  & 4.18 \\
 \quad year: 1979          & 400 & JTrk & 3.86 \\
                    & 480 & Trwi & 4.17 \\[5pt]
 \quad \textit{WSR-1}      & 50  & Js   & 4.10 \\
 \quad $\hat{T}_0 = 12.87$ & 105 & Jcu  & 3.96 \\
 \quad year: 1980          & 245 & Je   & 3.43 \\
                    & 320 & Jca  & 2.91 \\
                    & 455 & Jna  & 4.18 \\
                    & 515 & JTrk & 3.86 \\
                    & 575 & Trwi & 4.17 \\
\hline
\end{tabular*}
\end{table}

\section{Bayesian hierarchical model}\label{sec:Methods}\label{sec3}
\subsection{Physics based modeling}
\label{subsec:physics}
Following convention, we assume that for reduced temperatures,
boreholes are well approximated
as homogeneous, heat source free (except at the surface),
one-dimensional, semi-infinite solids
(i.e., a single boundary is at the surface).
It follows that the reduced temperatures, $T_r(z,t)$ at depth $z$ and
time $t$, can be reasonably modeled by the heat equation,
%
\begin{equation}
\label{eq:HeatEq}
\frac{\partial^2 T_r(z,t)}{\partial z^2} =
\frac{1}{\kappa} \,\frac{\partial T_r(z,t)}{\partial t},
\end{equation}
where $\kappa$ is the \textit{thermal diffusivity} of rock.
As is customary in
borehole analysis, we fix $\kappa$ to be $10^{-6}$ m$^2$/s
[see, e.g., \citet{HarrisChapman1995}].
The boundary condition, $T_r(0,t)$, is the primary target of
our inference.
Assuming that the initial reduced temperatures, $T_r(z,t_1)$,
are zero for all depths~$z$, the solution to (\ref{eq:HeatEq}) is
%
\begin{equation} \label{eq:sol1}
T_r(z,t) = \frac{2}{\sqrt{\pi}} \int_{z/2\sqrt{\kappa t}}^\infty
T_r \biggl(0, t - \frac{z^2}{4 \kappa\mu^2}  \biggr) e^{-\mu^2}\,d\mu
\end{equation}
[e.g., \citet{Carslaw1959}].
We assume that the boundary function is a step function,
%
\begin{equation} \label{eq:phi}
T_r(0,t) =
\cases{\displaystyle
T_1 ,&\quad   if   $t_1 < t < t_2 $,\cr\displaystyle
T_2 ,&\quad   if   $t_2 < t < t_3 $,\cr\displaystyle
\vdots\cr\displaystyle
T_{K} ,&\quad   if   $t_{K} < t < t_{K+1}$.
}
\end{equation}
Then the solution (\ref{eq:sol1}) reduces to
%
\begin{eqnarray}\label{eq:HESol}
T_r(z,t) & =& T_1 \mathrm{erfc} \biggl( \frac{z}{ \sqrt{4\kappa(t-t_1)}}
 \biggr)
+ (T_2-T_1) \mathrm{erfc} \biggl( \frac{z}{ \sqrt{4\kappa(t-t_2)}}
 \biggr)
\nonumber
\\[-8pt]
\\[-8pt]
&&{} + \cdots+ (T_{K}-T_{K-1}) \mathrm{erfc}
 \biggl( \frac{z}{ \sqrt{4\kappa(t-t_{K})}}  \biggr),
\nonumber
\end{eqnarray}
where $\mathrm{erfc}(\cdot)$ is the complementary error function
%
\begin{equation}
\mathrm{erfc}(x) = \frac{2}{\sqrt{\pi}} \int_x^\infty e^{-\mu^2}\,d\mu.
\end{equation}
Here, the time points $t_1, \ldots , t_{K+1}$ are
calendar years, $t_1$ being the earliest year considered
and $t_{K+1}$ is the year
the borehole data were collected. For example, borehole SRD-7 was
logged in $t_{12}=1980$ and we selected the following years for the
time points
$t_1, \ldots , t_{11}$: 1600, 1650, 1700, 1750, 1800, 1850, 1875, 1900,
1925, 1950 and 1965. This range is in concert with other analyses. We
also performed some analyses using a slightly more refined temporal
grid. These resulted in little change in the general form of the
posterior results. We note that for highly refined grids, the
structure and dimension of $A$ becomes an issue.

Collecting terms in (\ref{eq:HESol}), we can write the
solution at $t = t_{K+1}$ in vector form:
%
\begin{equation}
\label{eq:vectorsol}
\mathbf{T}_r = A \mathbf{T}_h,
\end{equation}
where $\mathbf{T}_r = (T_r(z_1,t_{K+1}),\ldots ,T_r(z_N,t_{K+1}))
^{\prime}$,
$\mathbf{T}_h = (T_1,\ldots ,T_{K})^{\prime}$, and $A$ is an
$N \times K$ matrix
with $(i,j)$th entry
\[
\mathrm{erfc} \biggl( \frac{z_i}{\sqrt{4 \kappa(t_{K+1} - t_{j})}}  \biggr)
- \mathrm{erfc} \biggl( \frac{z_i}{\sqrt{4 \kappa(t_{K+1} - t_{j+1})}}
 \biggr)
\]
for $i = 1,\ldots , N$ and $j = 1, \ldots , K-1$
and $(i,K)$th entry
\[
\mathrm{erfc} \biggl( \frac{z_i}{\sqrt{4 \kappa(t_{K+1} - t_{K})}}
 \biggr); \qquad i = 1,\ldots , N.
\]

\subsection{Single-site model} \label{sec:SingleSite}\label{sec3.2}
It is useful to view the modeling in
three basic stages, a data model, process model
and a parameter model.

(i) \textit{Data model}. Let $\mathbf{Y}$ be a vector of
observed temperatures at depths $z_1, \ldots , z_N$ and let
$\mathbf{T}$ denote the $N$-dimensional
vector of corresponding true
temperatures. We assume that the observations are
noisy, unbiased measurements of the true temperatures.
Specifically, we assume that
%
\begin{equation} \label{eq:YTtrue}
\mathbf{Y} = \mathbf{T} + \bepsilon,
\end{equation}
where $\bepsilon$ is an $N$-dimensional vector of normally
distributed, independent errors, all with mean zero and common
variance $\sigma_Y^2$ (see Section~\ref{sec:SensSpatial} for
discussion of the independence assumption).

Recalling the definition of reduced temperatures in (\ref{eq:Reduced}),
the true temperatures can be written as
%
\begin{equation} \label{eq:Tu}
\mathbf{T} = \mathbf{T}_r + T_0
\mathbf{1}_N + q_0 \mathbf{R},
\end{equation}
where $\mathbf{T}_r$ is the vector of true reduced temperatures and
other quantities are defined after (\ref{eq:Reduced}).

Combining (\ref{eq:YTtrue}) and (\ref{eq:Tu}), the assumed data model
is the
conditional distribution
%
\begin{equation}\label{eq:DataModel}
\mathbf{Y} | \mathbf{T}_r, q_0, \sigma_Y^2 \sim N(\mathbf{T}_r + T_0
\mathbf{1}_N + q_0 \mathbf{R}, \sigma_Y^2 I_N),
\end{equation}
where the vertical bar $ | $ is read ``given,'' $ \sim$ is read
``is distributed,'' and $I_N$ is the $N \times N$ identity matrix.

Note that $T_0$ and $\mathbf{T}_r$ in (\ref{eq:DataModel}) are not
\textit{identifiable} in that, for any constant~%
$c$, the shifted parameters
$\mathbf{T}_r \rightarrow \mathbf{T}_r + c\mathbf{1}_N$,
$T_0 \rightarrow T_0 - c$ yield identical data models.
To circumvent this issue, we assume that $T_0$ is known. The assumed
values of $T_0$ for the boreholes analyzed here are given
in Table~\ref{tab:Cond}. We report on the
sensitivity of results to the choice of $T_0$ in Section~\ref{sec:SensT0}.

(ii) \textit{Process model}.
We let $\mathbf{T}_h$ denote the $K$-vector containing the
surface temperature history.
We incorporate the heat equation in defining a~stochastic process model
%
\begin{equation}
\label{eq:ProcessModel}
\mathbf{T}_r | \mathbf{T}_h, \Sigma
\sim N(A \mathbf{T}_h, \Sigma)
\end{equation}
[recall (\ref{eq:vectorsol})].
We also assume a Gaussian prior for the histories:
%
\begin{equation}
\label{eq:HistoryPrior}
\mathbf{T}_h \sim N(\bmu, \Gamma).
\end{equation}

(iii) \textit{Parameter model}.
We assume that the covariance matrix of the process model errors
[see (\ref{eq:ProcessModel})] is diagonal with common variance,
namely, $\Sigma= \sigma^2 I_N$. That is, after accounting for the
dependence on the temperature history,
the heat equation offers
reliable explanation of reduced temperatures requiring only some
local-in-depth errors.
In Section~\ref{sec:SensSpatial} we consider sensitivity of
results with respect to the independence assumption.

Additional specifications of parameter priors is delayed until we
discuss modeling for multiple sites.

\subsection{Multiple-site model: Spatially distributed parameters}
\label{sec:MultSite}
We extend the model to combine data from multiple boreholes by
allowing site-specific processes and parameters.
A list of all the model parameters
used in this model is provided in Appendix \ref{app:List}.

(i) \textit{Data model}. We assume that measurements
from different sites are conditionally independent with data
models that depend only on site-specific processes and parameters. We
also assume that reduced temperature vectors from different sites are
conditionally independent with priors that depend on site-specific
histories and parameters. Formally, the data model is the product of
densities corresponding to the models
%
\begin{equation}
\label{eq:multipledatamodel}
\mathbf{Y}_j | \mathbf{T}_{r j}, q_{0 j}, \sigma_{Y_j}^2
\sim N_{N_j}(\mathbf{T}_{r j} + T_{0 j}
\mathbf{1}_{N_j} + q_{0 j} \mathbf{R}_j, \sigma_{Y_j}^2 I_{N_j})
\end{equation}
for the 9 sites labeled
$j=1,\ldots ,9$ with observation vectors of length $N_j$.

(ii) \textit{Process model}.
Similarly, the process model for the reduced temperature vectors is
the product of densities for
%
\begin{equation}
\label{eq:reduced_multipleprior}
\mathbf{T}_{r j} |
\mathbf{T}_{h j}, \sigma_j^2 \sim N_{N_j}(A_j \mathbf{T}_{h j},
\sigma_j^2 I_{N_j}).
\end{equation}

Since all 9 sites
are in the Colorado Plateau, we expect them to have been
influenced by common large-scale climate effects. However, the sites
are located in two subregions: Sites 1--5 are in the
San Rafael Desert (D), Sites~\mbox{6--9} are
in the San Rafael Swell (S).
To account for common influences within subregions,
we assume that all 9 histories are conditionally independent, with
parameters depending on subregions:
%
\begin{equation}
\label{eq:mulhistpriorD}
\mathbf{T}_{h j}| \bmu_D, \gamma_D^2
\sim N_{K}(\bmu_D, \gamma_D^2 I_{K}), \qquad
j = 1, \ldots , 5,
\end{equation}
and
%
\begin{equation}
\label{eq:mulhistpriorS}
\mathbf{T}_{h j} |\bmu_S, \gamma_S^2   \sim N_{K}(\bmu_S, \gamma_S^2
I_{K}), \qquad
j = 6, \ldots , 9.
\end{equation}
We remark that this is a very elementary spatial model. More complex
spatial modeling of parameters is feasible and recommended, depending
on prior information and data richness.

(iii) \textit{Parameter model}.

\textit{Model for heat flow parameters}:
To account for region-wide influences, we assume that the heat flow
parameters $\mathbf{q} = (q_{0 1}, \ldots , q_{0 9})^{\prime}$
are sampled from priors with dependence
structures. These priors are similar to \textit{exchangeable} models
[e.g., Section~4.6.2 in \citet{Berger1985}].

The heat flow parameters are assumed to be conditionally
independent with Gaussian priors where both
the mean and the variance depend on which region the borehole is in,
%
\begin{equation}
\label{eq:qpriorD}
q_{0 j} | \nu_D, \tau_D^2  \sim N(\nu_D, \tau_D^2), \qquad j = 1,
\ldots , 5,
\end{equation}
and
%
\begin{equation}
\label{eq:qpriorS}
q_{0 j} |\nu_S, \tau_S^2  \sim N(\nu_S, \tau_S^2), \qquad j = 6,
\ldots , 9.
\end{equation}
The means $\nu_D$ and $\nu_S$ are assumed to be a
priori independent with
Gaussian priors having common mean $\nu$:
%
\begin{equation}
\label{eq:nuDSprior}
\nu_D | \nu \sim N(\nu, \eta^2) \quad\mbox{and} \quad\nu_S |
\nu \sim N(\nu, \eta^2).
\end{equation}
Next, we assume that
%
\begin{equation}
\label{eq:nuprior}
\nu \sim N(\nu_0, \eta_0^2).
\end{equation}
We can combine the distributions in (\ref{eq:nuDSprior}) and
(\ref{eq:nuprior}) and integrate out $\nu$, leading to the
following prior for $\nu_D$ and $\nu_S$:
%
\begin{equation}
\label{eq:jointnu}
\pmatrix{\displaystyle \nu_D \cr\displaystyle  \nu_S
}
\sim
N \left(
\pmatrix{\displaystyle \nu_0 \cr\displaystyle  \nu_0
}
,
\pmatrix{\displaystyle \eta^2_D + \eta^2_0 & \eta^2_0 \cr\displaystyle
\eta^2_0 & \eta^2_S + \eta^2_0
}
\right ).
\end{equation}

\textit{Model for means of histories}:
We assume the means of the temperature histories $\bmu_D$ and
$\bmu_S$ in (\ref{eq:mulhistpriorD}) and (\ref{eq:mulhistpriorS})
have common mean $\bmu$; $\bmu_D|\bmu\sim N(\bmu, \sigma_D^2I_K)$ and
$\bmu_S|\bmu\sim N(\bmu, \sigma_S^2I_K)$. We in turn assume that
$\bmu\sim N(\bmu_0,\break \sigma_0^2 I_K)$. As in the development of
(\ref{eq:jointnu}), integrating out $\bmu$ yields
the following joint prior for $\bmu_D$ and $\bmu_S$:
%
\begin{equation}
\label{eq:historyjoint}
\pmatrix{\displaystyle \bmu_D \cr\displaystyle  \bmu_S
}
\sim
N \left(
\pmatrix{\displaystyle \bmu_0 \cr\displaystyle  \bmu_0
}
,
\pmatrix{\displaystyle (\sigma^2_D + \sigma^2_0) I_K & \sigma^2_0 I_K\cr\displaystyle
\sigma^2_0 I_K & (\sigma^2_S + \sigma^2_0) I_K
}
 \right).
\end{equation}
Note that the covariance structures in
(\ref{eq:mulhistpriorD}), (\ref{eq:mulhistpriorS}) and
(\ref{eq:historyjoint}) only include some spatial dependence, but no
temporal structure.
Some spatial-temporal dependence among historical
temperature values are displayed in their posterior distribution.

\textit{Priors for the variances}:
The measurement error and process model error variances appearing in
(\ref{eq:multipledatamodel})--(\ref{eq:qpriorS}) are all assigned
independent, inverse gamma priors:
%
\begin{eqnarray}\label{eq:PriorSigmas}
\sigma^2_{Y_j} & \sim& IG(a_Y, b_Y) \quad\mbox{and} \quad
\sigma_j^2 \sim IG(a,b)
\qquad\mbox{for } j = 1,\ldots , 9 ,\\
\tau_D^2 & \sim& IG(a_{\tau},b_{\tau}) \quad \mbox{and} \quad \tau_S^2 \sim IG(a_{\tau},b_{\tau}) , \\\label{eq:PriorGammas}
\gamma_D^2 & \sim& IG(a_{\gamma},b_{\gamma}) \quad\mbox{and} \quad
\gamma_S^2 \sim IG(a_{\gamma},b_{\gamma}) .
\end{eqnarray}

\subsection{Selection of parameters of prior distributions}
\label{sec:Priors}
We describe the selections of parameters of priors or \textit
{hyperparameters} introduced above:

\textit{Measurement error variances ($a_Y, b_Y$).}
As suggested in \citet{HarrisChapman1995}, ``The precision and accuracy
of the measurements are estimated to be
better than 0.01 K and 0.1 K, respectively.''
We view this as suggesting that a reasonable prior mean for the
variances of the
measurement errors,~$\sigma^2_{Y_j}$, $j=1,\ldots ,9$, is $0.11^2$. A
conservative choice for the prior variances is~$1.0$.
The values
$a_Y = 2.000146$ and $b_Y = 0.012102$ yield an inverse gamma
distribution matching these properties.
For additional intuition we remark
that the 0.025 and 0.975 quantiles of this prior are equal to
$0.002172$ and $0.049955$, respectively. Further, the corresponding
0.025 and 0.975
quantiles for the $\sigma_{Y_j}$ are $0.0466$ and $0.2235$, respectively.

\textit{Model error variances ($a,b$)}. Though we know comparatively
little about the variances $\sigma_j^2$
of the model errors, we can develop some plausible expectations.
For example, if the standard deviations of the
model errors are~0.5, we expect the model to be within $1.5^{\circ} C$
from the truth $99.7\%$ of the time. Hence, we specified the prior mean of
each $\sigma_j^2$ to be $0.50^2$ and a very large prior variance of
$100$. These selections correspond to
$a=2.000625$ and $b=0.250156$.
The corresponding 0.025 and 0.975
quantiles for the $\sigma_{j}$ are $0.212$ and $1.016$, respectively.

\textit{Heat flow parameters ($\nu_0,\eta^2_D,\eta^2_S,
\eta^2_0,a_{\tau}, b_{\tau}$)}.
The background heat flow $q_0$ has been shown in other studies to
range from about 30 mW/m$^2$ (milliwatt per meter-squared) to
about 100 mW/m$^2$, with the majority of values ranging between 50 and
70 mW/m$^2$
[\citet{BodellChapman1982}; \citet{BeltramiMareschal1995}; \citet{DorofeevaShenShapova2002}, e.g.].
Focusing on (\ref{eq:jointnu}),
we selected the prior mean $\nu_0 = 60$ mW/m$^2$ and set the
standard deviation $\eta_0 = 20$ mW/m$^2$.
The standard deviations $\eta_D$ and $\eta_S$
represent variability due to subregion. We set
$\eta_D = \eta_S = 10$. Note that these selections imply that
the prior standard deviations of $\nu_D$ and $\nu_S$ are equal to
$(20^2 +
10^2)^{ 0.50} \approx22.36$ mW/m$^2$ and the correlation between
$\nu_D$ and
$\nu_S$ is~$0.80$. We discuss sensitivities of results to these
selections in Section~\ref{sec:SensEta}.

Recalling (\ref{eq:qpriorD}) and (\ref{eq:qpriorS}),
$\tau_D^2$ and $\tau_S^2$ quantify variability of
the $q_{0 j}$ about their regionally defined prior means. We set the
prior means for these variances to $0.1^2$ with a corresponding large prior
variance of 1. It follows that we select
$a_{\tau}=2.000100$ and $b_{\tau}= 0.010001$. Note that the units here
are W/m$^2$, so
this corresponds to $\tau^2_D$ and $\tau^2_S$ having prior mean of
$100^2$ (mW/m$^2$)$^2$
and the 0.025 and 0.975 quantiles for
$\tau_{D}$ and $\tau_S$ are
$42.4$ mW/m$^2$ and $203.2$~mW/m$^2$, respectively.

\textit{Histories} ($\bmu_0,
\sigma^2_D,\sigma^2_S,\sigma^2_0,a_{\gamma}, b_{\gamma}$).
In the model parameterizations used here both reduced temperatures and
temperature histories represent departures from the baseline surface
temperature $T_0$ [see (\ref{eq:Tu})]. Hence, a reasonable prior mean
for $\mathbf{T}_h$ is $\bmu_0 = \mathbf{0}$. Further, these departures
occur over time intervals of lengths between 10 and 50 years. Focusing
on (\ref{eq:historyjoint}), we selected $\sigma^2_0 = 0.1$ and
$\sigma^2_D = \sigma^2_S = 0.2$. These selections imply that
the prior standard deviations of the coordinated
$\mu_{D k}$ and $\mu_{S k}$, $k=1, \ldots , K$, are equal to $(0.1
+0.2)^{ 0.50} \approx 0.5477$
and the correlation between $\mu_{Dk}$ and
$\mu_{Sk}$ is $0.333$. We discuss sensitivities of results to these
selections in Section \ref{sec:SensSigma}.

Finally, recalling (\ref{eq:mulhistpriorD}) and (\ref{eq:mulhistpriorS}),
$\gamma_D^2$ and $\gamma_D^2$ quantify variability of
the elements of the history vectors
about their regionally defined prior means. We set the
prior means for these variances to $0.8$ and prior variances
equal to 1. It follows that we select
$a_{\gamma}= 2.064$ and $b_{\gamma} = 0.8512$ and the corresponding
0.025 and 0.975
quantiles for $\gamma_{D}$ and $\gamma_s$ are $0.387$ and $1.801$, respectively.

\section{Results} \label{sec:Results}\label{sec4}

\subsection{Multiple-site model} \label{sec:ResultsMulti}
$\!\!\!$The multiple-site model described in Section~\ref{sec:MultSite}
has 818 unknown parameters, including 664 elements of
the reduced temperature\vadjust{\goodbreak} vectors $\mathbf{T}_{rj}$.
We implemented a Gibbs sampler in \texttt{R} to
obtain samples from the posterior distribution.
The full conditional distributions used are given in Appendix~\ref{app:Gibbs}.
We obtained 30,000 samples and discarded the first 2000 as burn-in,
leaving 28,000 samples to use for inference.
Trace-plots for parameters and a random selection of elements of
$\mathbf{T}_{rj}$
showed no indication of convergence problems.
We estimated marginal posterior densities
using (Gaussian) kernel density estimation.

\begin{figure}[t]

\includegraphics{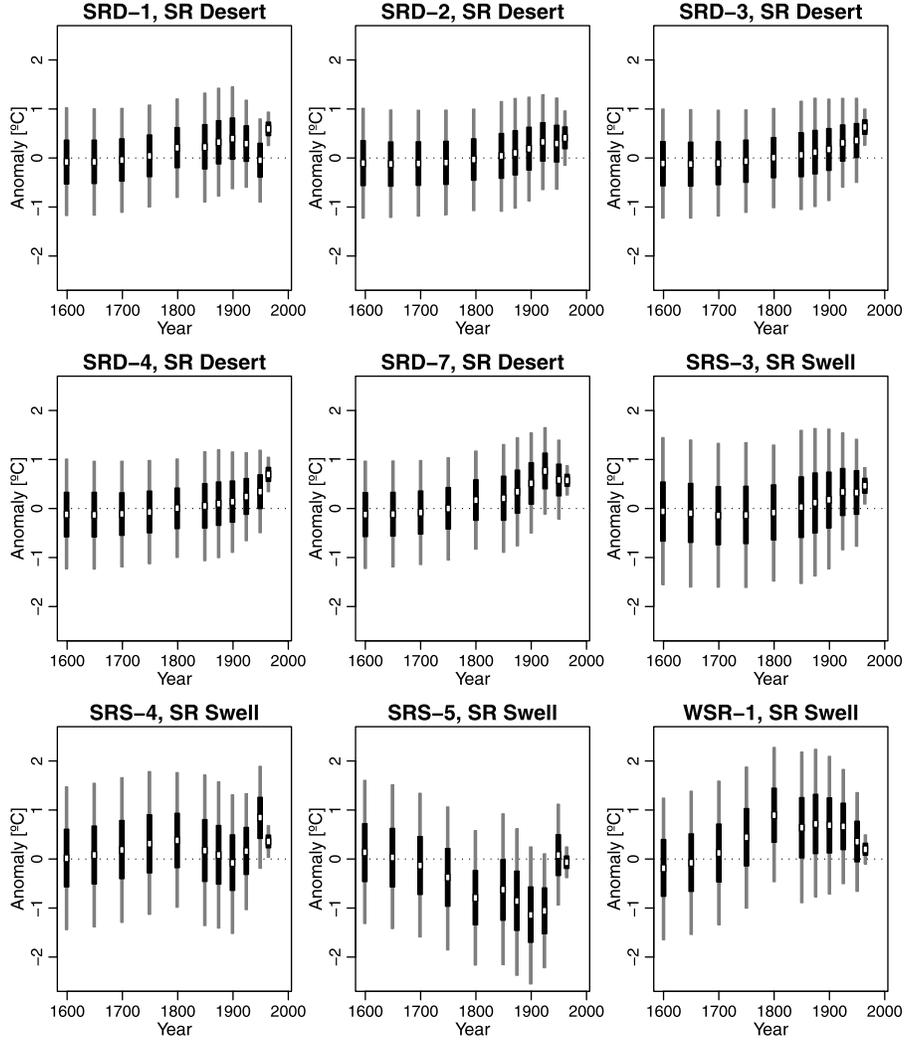}
\vspace*{-5pt}
\caption{Estimated posterior means and credible sets for the
ground surface temperature (GST) histories, $\mathbf{T}_{hj}$.
The white squares show the posterior means of $\mathbf{T}_{hj}$
and the vertical bars show symmetric 50\% (thicker and black) and
90\% (thinner and grey) posterior credible intervals.}\label{fig:GTH}
\vspace*{-7pt}
\end{figure}

\begin{figure}

\includegraphics{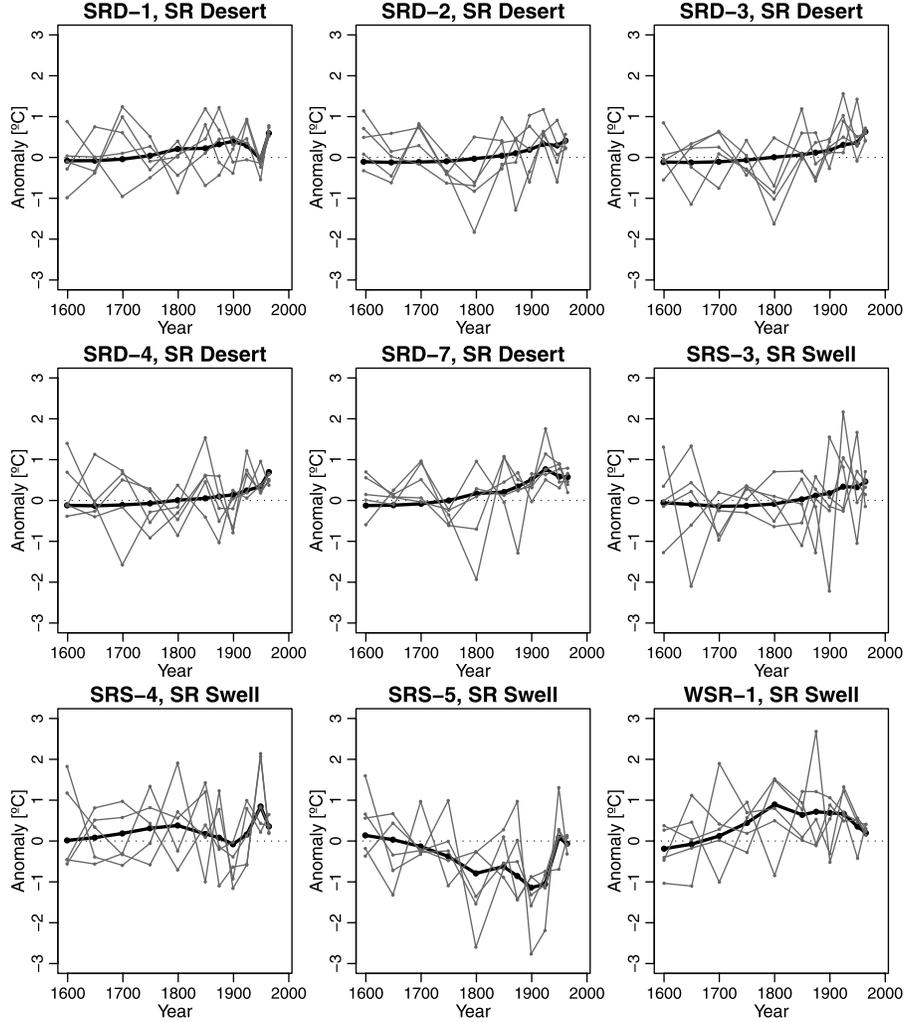}
\vspace*{-5pt}
\caption{Ensembles of ground surface
temperature (GST) histories generated from the posterior distribution.
For each borehole, the thick black lines show the posterior means of
$\mathbf{T}_{hj}$ and the five
grey lines are five different samples of
$\mathbf{T}_{hj}$.}\label{fig:GTHEnsembles}
\vspace*{-7pt}
\end{figure}

\begin{figure}

\includegraphics{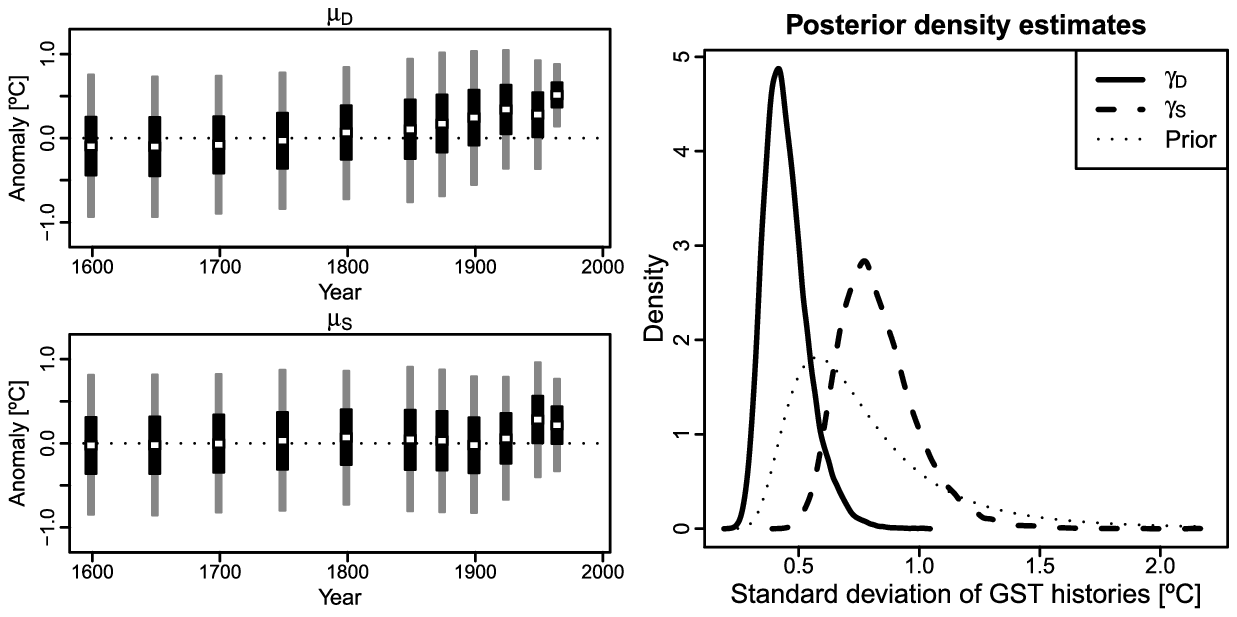}

\caption{\textup{Left:} Estimated posterior means (white squares)
and symmetric 50\% and 90\%
posterior credible intervals for the mean GST
histories $\bmu_D$ \textup{(upper)} and $\bmu_S$ \textup{(lower)}.
\textup{Right:} Estimated posterior densities for the standard deviations of the
GST histories for both areas, $\gamma_D$ and $\gamma_S$.
The prior density is the same
for both $\gamma_D$ and $\gamma_S$ (dotted line).}\label{fig:MuGamma}\vspace*{-2pt}
\end{figure}

The ground surface temperature (GST) histories,
$\mathbf{T}_{hj}$, are of primary interest.
Estimated posterior means and credible sets for the GST histories
are shown in Figure \ref{fig:GTH}.
In Figure \ref{fig:GTHEnsembles} we show 5 samples of
$\mathbf{T}_{hj}$ for each borehole. These 5 samples
are taken 6000 MCMC iterations apart
(after burn-in) and 100 iterations apart between boreholes. Note that
the generated realizations are not overly smooth, but that the
spreads of the realizations are decreasing as time approaches the
present.
 Estimated posterior means and
credible sets
for the mean GST histories for the San Rafael (SR) Desert and SR Swell
regions, $\bmu_D$, $\bmu_S$,
and prior and posterior densities of the parameters $\gamma_D$
and $\gamma_S$, are shown in Figure \ref{fig:MuGamma}.
One striking feature of the GST histories is that posterior
uncertainties are substantial.
We see patterns of warming in the last century
(and cooling for site WSR-1),
but there are large posterior uncertainties associated with the estimated
trends at each borehole.
On the other hand, the posterior uncertainty is lower for more
recent times (especially at the last time point).

We note that the temperature trends (posterior means) are similar
within the
SR Desert sites, but quite variable
within the SR Swell region, most notably at sites SRS-5 and WSR-1.
Furthermore, the posterior
credible intervals are wider for boreholes in the SR Swell.
This difference is also apparent in the density estimates of the
standard deviation of the GST histories:
$\gamma_S$ is larger than $\gamma_D$ (see Figure \ref{fig:MuGamma},
right). The mean GST histories
${\bmu}_D$ and ${\bmu}_S$ show slightly different temperature trends
(Figure \ref{fig:MuGamma}, left),
but the posterior uncertainty is somewhat large and
increasing as we go further back in time.

\begin{figure}[b]

\includegraphics{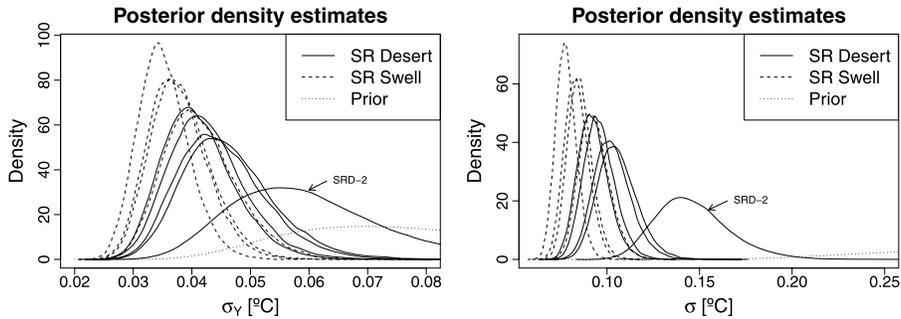}

\caption{Estimated posterior
densities of the measurement error standard deviations
$\sigma_{Yj}$ \textup{(left)} and the model error
standard deviations $\sigma_j$ \textup{(right)}. In both cases the prior
is the same for all nine boreholes (dotted lines).}\label{fig:SigmaAll}\vspace*{-2pt}
\end{figure}

Estimated marginal posterior densities of the site-wise measurement and model
error standard deviations,
$\sigma_{Yj}$ and $\sigma_j$, are shown in Figure \ref{fig:SigmaAll}.
The locations of these densities are quite different from the prior means.
The~$\sigma_{Yj}$ are of the order $0.03$--$0.05^\circ$C
compared to their prior mean $0.11^\circ$C. The $\sigma_j$ are of the
order $0.05$--$0.15^\circ$C compared to the prior mean $0.5^\circ$C.
An interesting pattern emerges in Figure \ref{fig:SigmaAll}.
Both $\sigma_{Yj}$ and $\sigma_j$ have higher posterior uncertainty
for boreholes in the SR Desert than the SR Swell.
Note that the boreholes in the SR Swell have more measurements
than those in the SR Desert (see Figure \ref{fig:Data}).
The SRD-2 borehole has by far the fewest measurements and,
as indicated in Figure \ref{fig:SigmaAll}, densities
for $\sigma_{Yj}$ and $\sigma_j$ for that site are wider and
are closer to the prior than
the other densities.

\begin{figure}

\includegraphics{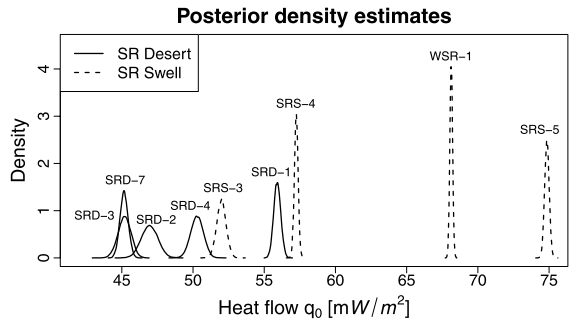}

\caption{Estimated posterior densities of the heat
flow $q_{0j}$ for boreholes from the San Rafael Desert (solid lines)
and San Rafael Swell (long dashes).}\label{fig:DensQ0}
\end{figure}

\begin{figure}[t]

\includegraphics{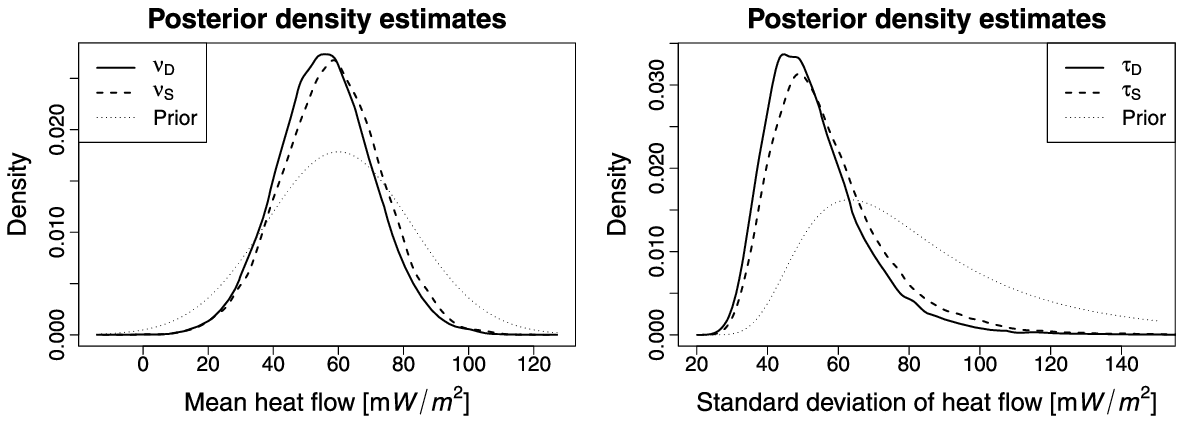}

\caption{Estimated posterior densities of the
means, $\nu_D$ and $\nu_S$ \textup{(left)},
and the standard deviations, $\tau_D$ and $\tau_S$ \textup{(right)},
of the background heat flow for both regions.
The (marginal) prior distributions are the same for both regions and are
presented with dotted lines.}\label{fig:DensNuTau}
\end{figure}

Estimated posterior densities of the background heat flow $q_{0j}$
are shown in Figure
\ref{fig:DensQ0}. Posterior means and 90\% credible intervals
for $q_{0j}$ and the means $\nu_D$ and $\nu_S$
are shown in Table \ref{tab:Q0Nu}.
Estimated posterior densities of the means and standard deviations of
the heat flow, $\nu_D$, $\nu_S$, $\tau_D$ and $\tau_S$,
are shown in Figure \ref{fig:DensNuTau}.
It is clear from Figure \ref{fig:DensQ0} that
the background heat flow is lower for boreholes in the SR
Desert than in the SR Swell (except for sites \mbox{SRD-1} and SRS-3).
The 9 posterior\vadjust{\goodbreak} densities show varying degrees of posterior
uncertainty, in large part in response
to the amount of data in each borehole.
The high values of the standard deviations $\tau_D$ and $\tau_S$
(see Figure~\ref{fig:DensNuTau})
indicate the wide range of the heat flow $q_{0j}$ within the regions.

\begin{table}[b]
\tabcolsep=0pt
\caption{Posterior means and
symmetric 90\% credible intervals (CI) for the background heat flows
$q_{0j}$ (mW/m$^2$) and the mean heat flows $\nu_D$ and $\nu
_S$ (mW/m$^2$)}\label{tab:Q0Nu}
\begin{tabular*}{\textwidth}{@{\extracolsep{\fill}}lccccc@{}}
\hline
\multicolumn{3}{@{}c}{\textbf{San Rafael Desert}} & \multicolumn{3}{c@{}}{\textbf{San Rafael Swell}} \\[-5pt]
\multicolumn{3}{@{}c}{\hrulefill} & \multicolumn{3}{c@{}}{\hrulefill}\\
\textbf{Borehole} & \textbf{Mean}& \textbf{90\% CI} & \textbf{Borehole} & \textbf{Mean} & \textbf{90\% CI} \\
\hline
SRD-1 & 55.91 & (55.51, 56.32) & SRS-3 & 51.99 & (51.45, 52.54) \\
SRD-2 & 46.95 & (46.02, 47.90) & SRS-4 & 57.25 & (57.02, 57.47) \\
SRD-3 & 45.19 & (44.42, 45.92) & SRS-5 & 74.85 & (74.58, 75.11) \\
SRD-4 & 50.28 & (49.54, 51.01) & WSR-1 & 68.13 & (67.97, 68.29) \\
SRD-7 & 45.16 & (44.69, 45.63) & & & \\ [3pt]
$\nu_D$ & 55.95 & (31.94, 80.39) & $\nu_S$ & 58.05 & (32.93, 83.08) \\
\hline
\end{tabular*}
\end{table}

There is considerable posterior uncertainty regarding the
mean heat flows $\nu_D$ and $\nu_S$ (see
Figure \ref{fig:DensNuTau}), especially when compared to the
relatively precise posterior densities for the 9 individual
heat flows $q_{0j}$. This is what we expect since there is
substantial variation of the locations of these precise,  individual densities.
Another point is that the posterior means of
$\nu_D$ and~$\nu_S$ are surprisingly alike, and do not seem to
correspond to
the difference in heat flows for the two regions that is apparent in
Figure \ref{fig:DensQ0}.
For example, the posterior mean of $\nu_D$
is higher than all the posterior means of $q_{0j}$
in the SR Desert (see Table \ref{tab:Q0Nu}).
We explain this behavior in Section \ref{sec:SensEta}.

\subsection{Comparison to single-site models}\label{sec4.2}
\label{sec:ResultsSingle}
In our primary analysis we combined data from all boreholes in one hierarchical
multiple-site model.  The temperature histories for boreholes in the same
region share the same mean and variance. Similarly, the heat flow parameters
for boreholes within the same region share the same mean and variance.
One rationale for doing this is that it enables us to learn about region-wide
mean temperature histories and region-wide mean heat flow. Another rationale
is that hierarchically linking boreholes within regions allows for
sharing of information
between boreholes through the shared parameters.
This sharing of information across groups of data is often called
\textit{borrowing strength} and can often lead to better parameter estimates.
However, the question here becomes how much strength, if any,
is borrowed between the nine boreholes.
To assess this aspect of the model, we fit single-site models
described in Section \ref{sec:SingleSite} to each of the 9 boreholes.

By performing separate single-site
models, we of course do not model parameters as spatially dependent.
Operationally, some parameters treated as random in the combined
analysis are assigned fixed values. For example, we assign values
to $\bmu$ and $\Gamma$ in the prior for the GST histories
[see (\ref{eq:HistoryPrior})], as compared to the additional
stage involving $\bmu_D$, $\bmu_S$, $\gamma^2_D$
and $\gamma^2_S$ [see (\ref{eq:historyjoint}) and
(\ref{eq:PriorGammas})] in the combined analysis.
To make the results comparable, we set the prior
mean and covariance matrix of~the GST histories
equal to their marginal prior mean and covariance
implied by the multiple-site model.
For boreholes in the SR Desert ($j=1,\ldots ,5$), we have the following:
%
\begin{eqnarray}
E(\bT_{hj}) & =& E( E(\bT_{hj} | \bmu_D) ) = E( \bmu_D) = \bmu_0 =
\mathbf{0}, \\
\operatorname{Cov}(\bT_{hj}) & =& \operatorname{Cov}( E(\bT_{hj} | \bmu_D) ) +
E( \operatorname{Cov}(\bT_{hj} | \bmu_D) )\nonumber \\
& =& \operatorname{Cov}( \bmu_D ) + E(\gamma^2_D I_K ) =  ( \sigma^2_D + \sigma^2_0
 ) I_K + E(\gamma^2_D) I_K \\
& =& (0.2 + 0.1 +0.8) I_K = 1.1 I_K . \nonumber
\end{eqnarray}
For boreholes in the SR Swell ($j=6,\ldots ,9$),
we also have $E(\bT_{hj}) = E( \bmu_S) = \mathbf{0}$ and
$\operatorname{Cov}(\bT_{hj}) = 1.1 I_K$.
Hence, we set the following prior
for $\bT_h$ in the single-site models (same for every borehole):
%
\begin{equation}
\bT_h \sim N(\bmu, \Gamma) = N_K(\mathbf{0}, 1.1 I_K).
\end{equation}
Similarly, we set the following prior for each $q_{0}$:
%
\begin{equation}
q_{0} \sim N( 0.06, 0.02^2 + 0.01^2 +0.1^2 = 0.0105).
\end{equation}
Finally, we used the same priors for measurement and model error
variances [see (\ref{eq:DataModel})
and (\ref{eq:ProcessModel})] as for the multiple-site model.

We fitted the single-site models via Gibbs samplers,
obtaining 10,000 MCMC samples from the posterior
distribution for each borehole and deleted 2000 iterations for burn-in.

\begin{figure}

\includegraphics{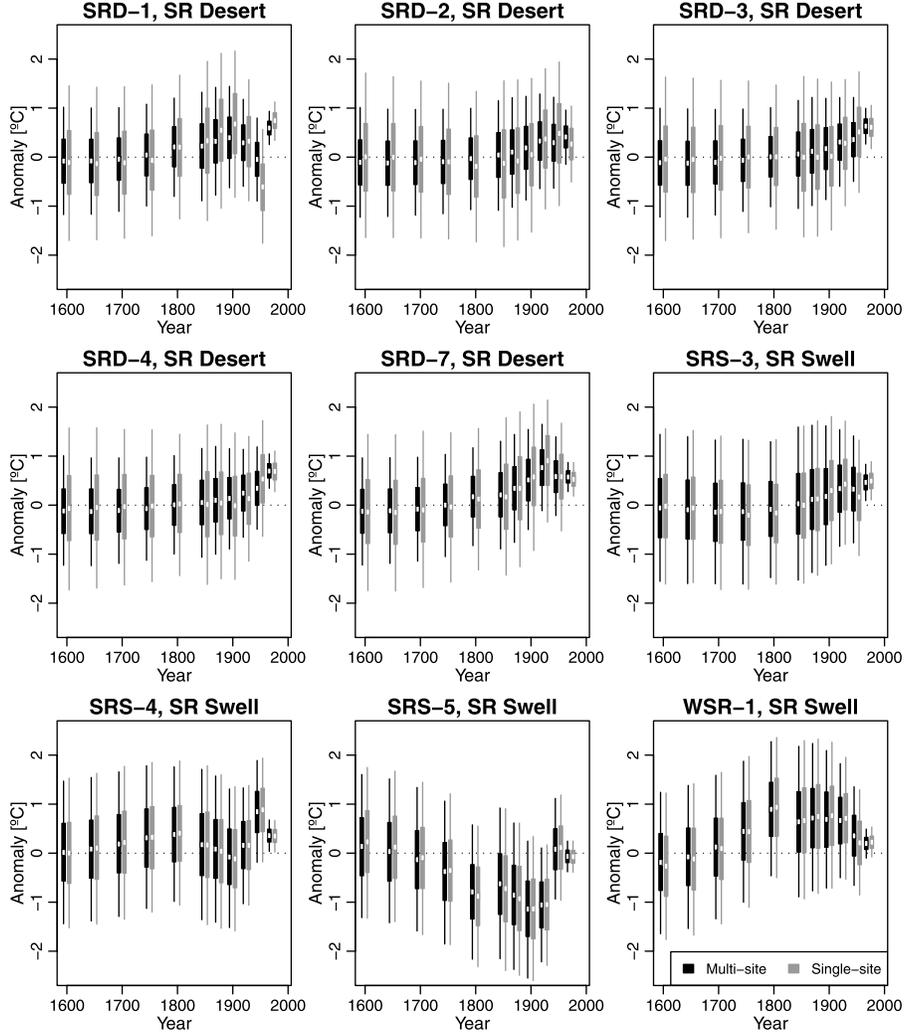}

\caption{Comparison of posterior distributions of
GST histories
from the multiple-site model (black) and single-site models (grey).
The white squares show the estimated posterior means and the vertical
bars show the
symmetric 50\% (thick) and 90\% (thin) posterior credible intervals.}\label{fig:GTHsingle}
\end{figure}

The estimated posterior means and credible sets for the GST histories from
both the multiple-site and the 9 single-site models are shown in
Figure~\ref{fig:GTHsingle}.
Focusing on the five boreholes in the SR Desert,
we see two major differences.
First, the GST posterior means (white squares in Figure \ref{fig:GTHsingle})
are slightly more dampened for the multiple-site model than the
single-site models. In other words,
we see shrinkage in the posterior means when we combine the boreholes.
Second, as indicated
by narrower 50\% and 90\% posterior credible intervals, the posterior
uncertainty is substantially less for the multiple-site model
than the single-site models.
We conclude that by combining the boreholes in the SR
Desert, the GST history parameters are borrowing
strength across boreholes.

In the SR Swell the posterior results are quite similar
for the multiple-site model and the single-site models.
In particular, the posterior uncertainties are
similar in that the 50\% and 90\% posterior credible intervals
are only slightly wider for the single-site models than
the multiple-site model.
We believe that the reason for these different results for the
two regions is the following: The SR Desert GST
histories are comparatively similar, so the analysis amplifies\vadjust{\goodbreak}
combining and borrowing strength. The SR
Swell results seem more diverse, suggesting borrowing strength should
be comparatively weak.

Posterior density estimates
(not shown here) for the background heat flows $q_{0j}$
and the variances $\sigma^2_Y$ and $\sigma^2$ were almost identical
for the two analyses.

\subsection{Sensitivity analyses} \label{sec:Sens}\label{sec4.3}
In our analyses we used prespecified fixed values of the
temperature intercepts $T_{0j}$ and hyperparameters
(i.e., fixed parameters of prior distributions).
The selection of hyperparameters was discussed in Section
\ref{sec:Priors}. We next assess the
sensitivity of results to some of these specifications.

\subsubsection{Temperature intercept $T_{0j}$} \label{sec:SensT0}
The temperature intercepts $T_{0j}$ used here were
least square estimates of the intercepts in simple
linear regressions. That is, separately for each borehole,
temperature data were regressed on the thermal resistance
vector $\mathbf{R}_j$. In these steps, the regressions were based only on
data at the deep parts of the borehole, specifically below 150 meters
for boreholes in the SR Desert and below 200 meters for
boreholes in the SR Swell.
We used the usual least squares
standard errors to guide our sensitivity analysis on $T_{0j}$.
Specifically, we fitted the multiple-site model in two additional
cases: (1) all $T_{0j}$'s were set to three standard errors
below the least squares estimates and (2) all $T_{0j}$'s set to
three standard errors above the least squares estimates.

Overall the results were not highly sensitive to these changes in
the temperature intercepts.
Estimated densities of the heat flow parameters $q_{0j}$ (not shown
here) indicated that posterior for the $q_{0j}$ responded to
changes in the $T_{0j}$'s as we expect a slope to change when the
intercept is changed.
When the $T_{0j}$ were lowered, the $q_{0j}$ were higher and vice versa.
However, posterior means and standard deviations
of the heat flows, $\nu_D$, $\nu_S$, $\tau_D$, $\tau_S$,
and $\sigma_{Yj}$ and $\sigma_j$, did not vary much as we
changed $T_{0j}$'s.

\begin{figure}

\includegraphics{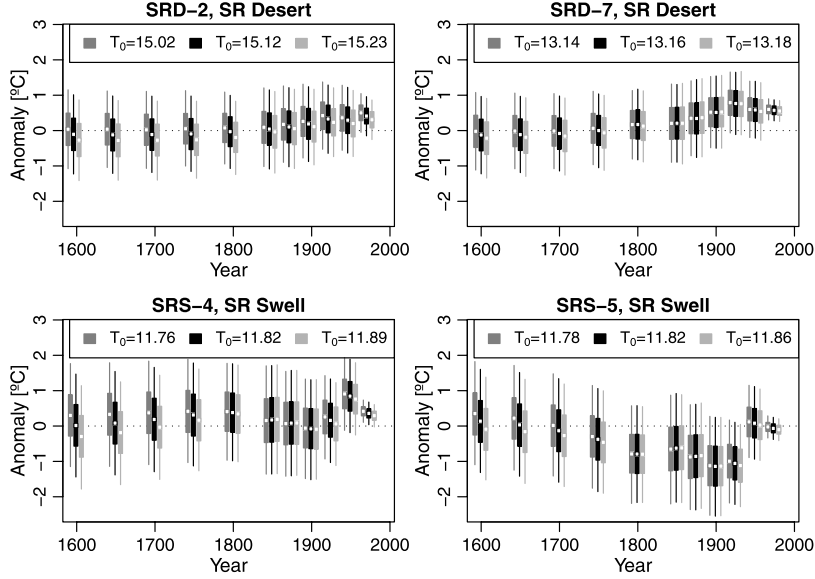}

\caption{Comparison of posterior distributions of
GST histories for
4 boreholes using different values for the temperature intercept $T_{0j}$.
The white squares show the estimated posterior means and the vertical bars
show the symmetric 50\% (thick) and 90\% (thin) posterior credible
intervals. At each time point the middle bar
shows the original results (same as in Figure \protect\ref{fig:GTH})
and the left and right bars show the results
for lower and higher values of $T_{0j}$, respectively.}\label{fig:GTHsensT0}
\end{figure}

The estimated posterior means and credible sets for the GST histories
for the three cases of $T_{0j}$ are
shown in Figure
\ref{fig:GTHsensT0}. We show results for two boreholes from each region,
but the effects are similar for the other boreholes.
The main effect of changing the
$T_{0j}$'s is that the posterior means of the
GST histories are slightly shifted. The temperature anomalies are
higher when $T_{0j}$ is lower and vice versa,
so it seems that they are compensating the changing~$T_{0j}$.
Interestingly, the amount of shifting increases as we go further back
in time.
On the other hand, the changes in posterior means are
not large when compared to the posterior uncertainty of
the results.
We conclude that sensitivities to the specifications of $T_{0j}$
are overshadowed by the posterior uncertainty of the results.

\subsubsection{Hyperparameters $\eta_D$, $\eta_S$, $\eta_0$}
\label{sec:SensEta}
Recalling (\ref{eq:jointnu}), the regional prior means,
$\nu_D$ and $\nu_S$, of the heat flows have prior standard deviations
$(\eta^2_D + \eta^2_0)^{0.5}$
and $(\eta^2_S + \eta^2_0)^{0.5}$, respectively,
and prior correlation
$\eta^2_0 /( (\eta^2_0 + \eta^2_D)(\eta^2_0 + \eta^2_S) )^{0.5}$.

We report on results for four additional settings of
$\eta_0$, $\eta_D$ and $\eta_S$ leading to
the prior standard deviations and correlations given in
Table \ref{tab:SensEta}.
Note that in light of the range of heat flow estimates
found in the literature (see Section \ref{sec:Priors}),
a prior standard deviation of 120 mW/m$^2$ is very large.
Also, the prior correlation we selected for the original setting is
very high (0.8).
Note also that sensitivity settings 2 and 4 have the same correlation
but different standard deviations.

\begin{table}
\tabcolsep=0pt
\caption{The original and four new settings
of the hyperparameters $\eta^2_D$, $\eta^2_S$ and $\eta^2_0$ along
with the implied prior standard deviations
of $\nu_D$ and $\nu_S$ and the prior correlations
between $\nu_D$ and $\nu_S$}\label{tab:SensEta}
\begin{tabular*}{\textwidth}{@{\extracolsep{\fill}}lcccc@{}}
\hline
& $\bolds{\eta^2_D}$ \textbf{and} $\bolds{\eta^2_S}$ & $\bolds{\eta^2_0}$
& \textbf{Prior sd. [mW/m$^{\bolds{2}}$]}& \textbf{Prior
cor} \\
\hline
Original setting & \hphantom{1}$10^2$ & $20^2$ & \hphantom{1}22.36 & 0.80 \\ [3pt]
Setting 1 & \hphantom{1}$20^2$ & $20^2$ & \hphantom{1}28.28 & 0.50 \\
Setting 2 & \hphantom{1}$30^2$ & $20^2$ & \hphantom{1}36.06 & 0.31 \\
Setting 3 & $100^2$ & \hphantom{6}$0^2$ & 100.00 & 0.00 \\
Setting 4 & $100^2$ & $67^2$ & 120.37& 0.31 \\
\hline
\end{tabular*}
\end{table}

First, posterior means and credible intervals of the GST histories
($\mathbf{T}_{hj}$; not shown) were almost identical across all five
cases. The same was true for the means and standard deviations of the
histories ($\bmu_D$, $\bmu_D$, $\gamma_D$ and $\gamma_S$; not shown)
Also, the posterior density estimates (not shown) for heat
flow parameters $q_{j}$ were almost identical in all cases.

\begin{figure}

\includegraphics{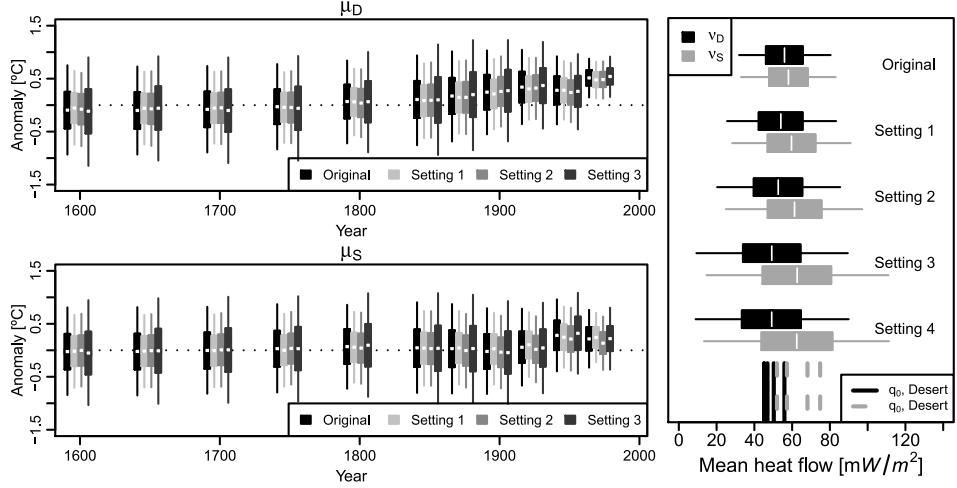}

\caption{\textup{Left:} Comparison of posterior
distributions of the mean
GST histories, $\bmu_D$ and~$\bmu_S$, using
different values of $\sigma^2_D$, $\sigma^2_S$ and $\sigma^2_0$
(see Table \protect\ref{tab:SensSigma}).
The white squares show the estimated posterior means and the vertical
bars show the
symmetric 50\% (thick) and 90\% (thin) posterior credible intervals.
\textup{Right:} Comparison of
mean heat flow $\nu_D$ and $\nu_S$
using different values of $\eta^2_D$, $\eta^2_S$ and $\eta^2_0$ (see
Table \protect\ref{tab:SensEta}).
The white squares show the estimated posterior means and the vertical
bars show the
symmetric 50\% (thick) and 90\% (thin) posterior credible intervals.
The line segments at the bottom show
the posterior means of the heat flows $q_{0j}$.}\label{fig:MuSensSigma}
\end{figure}

The posterior distributions of
$\nu_D$ and $\nu_S$ displayed strong sensitivities.
Figure~\ref{fig:MuSensSigma} (right) shows the posterior means
and credible intervals for $\nu_D$ and~$\nu_S$ for all 5 settings.
Not surprisingly, as the prior uncertainty increases,
the posterior uncertainty increases.
We note differences in the posterior means that correspond
to the different prior correlations. In the original setting
the posterior means of $\nu_D$ and $\nu_S$
were very close and seemed not to respond to the regional information
indicated in the posteriors of the $q_{0j}$ [the vertical line
segments in Figure~\ref{fig:MuSensSigma} (right) are the posterior
means of the~$q_{0j}$; also see Table \ref{tab:Q0Nu}].
Note that as the prior correlation decreases,
the posterior means of $\nu_D$ and $\nu_S$ separate
and approach the averages of the heat flow posterior means
in their corresponding regions.
However, due to the relatively large posterior
uncertainties of $\nu_D$ and $\nu_S$ in all cases, the changes
we see in the posterior means are all within the
50\% credible interval of the original model.

\begin{table}[b]
\tabcolsep=0pt
\caption{The original setting of the
hyperparameters $\sigma^2_D$, $\sigma^2_S$ and $\sigma^2_0$
and the three additional settings along with the
implied prior standard deviations
of the elements of the vectors $\bmu_D$ and $\bmu_S$ and~the~prior~correlations between $\mu_{Dk}$ and $\mu_{Sk}$
for $k=1, \ldots , K$}\label{tab:SensSigma}
\begin{tabular*}{\textwidth}{@{\extracolsep{\fill}}lcccc@{}}
\hline
& $\bolds{\sigma^2_D}$ \textbf{and} $\bolds{\sigma^2_S}$ & $\bolds{\sigma^2_0}$ & \textbf{Prior sd. (}$\bolds{{}^\circ
}$\textbf{C)}& \textbf{Prior cor} \\
\hline
Original setting & 0.2 &  0.1\hphantom{5} & 0.55 & 0.33 \\
[3pt]
Setting 1 & 0.1 &  0.1\hphantom{5} & 0.45 & 0.50 \\
Setting 2 & 0.2 &  0.0\hphantom{5} & 0.45 & 0.00 \\
Setting 3 & 0.3 & 0.15 & 0.67 & 0.33\\
\hline
\end{tabular*}
\end{table}

While these results are not surprising, it is instructive to see
the workings of Bayesian updating of borrowing-strength priors.
Finally, though the
sample sizes at each borehole are relatively large, the operative
``sample sizes'' for treating $\nu_D$ and $\nu_S$ are 5 and
4 sites, respectively.

\subsubsection{Hyperparameters $\sigma_D$, $\sigma_S$, $\sigma_0$}
\label{sec:SensSigma}
The hyperparameters $\sigma_D$, $\sigma_S$ and~$\sigma_0$ determine
the prior standard deviations, $(\sigma^2_D + \sigma^2_0)^{0.5}$ and
$(\sigma^2_S + \sigma^2_0)^{0.5}$, of the elements of the prior mean
GST history vectors $\bmu_D$ and $\bmu_S$.
They also imply that the prior correlations
$\operatorname{corr}(\mu_{Dk},\mu_{Sk})$, $k=1, \ldots , K$, are equal to
$\sigma^2_0 /( (\sigma^2_0 + \sigma^2_D)(\sigma^2_0 + \sigma^2_S) )^{0.5}$.
We considered three additional settings for these parameters
as shown in Table~\ref{tab:SensSigma}.

Overall, the results were not sensitive to the different settings of
$\sigma_D$, $\sigma_S$ and $\sigma_0$.
The mean GST histories in each region, $\bmu_D$ and $\bmu_S$, were
only slightly affected by the changes
in the prior [see Figure~\ref{fig:MuSensSigma} (left)].
The posterior credible intervals are slightly wider when the prior
standard deviations are higher. The posterior means are almost
identical for the four settings over the first three or four centuries.
In the last century there is a slight difference between settings 1 and 2,
particularly in the SR Swell. These settings
have the same prior standard deviation but different prior
correlations (0.5 and 0, resp.). It seems that
higher prior correlation leads to posteriors that favor
similarity of the elements
of $\bmu_D$ and $\bmu_S$, but only for the more
recent time points.
Again, we note that these differences are very small compared to the posterior
uncertainty in the results.

The results for the GST histories ($\bT_{hj}$, not shown here) were
very similar
for all four settings. The only differences were
that the posterior uncertainties increased slightly
when the prior standard deviations of $\bmu_D$ and $\bmu_S$
increased. The different prior
correlations did not seem to matter,
settings 1 and 2 gave almost identical results.
Density estimates (not shown) for other parameters
were virtually identical for the different prior settings.

\subsubsection{Correlated measurement and model errors}
\label{sec:SensSpatial}

Though we expect substantial correlation among the elements of each
$\mathbf{Y}_j$ and among the elements of each
$\mathbf{T}_{r j}$, we assumed conditionally independent measurement
errors and model errors in (\ref{eq:multipledatamodel}) and
(\ref{eq:reduced_multipleprior}), respectively. The modeling notions
are that the lion's shares of these correlations are due to the structure
of the true temperatures in the case of~the $\mathbf{Y}_j$ and the
structure in temperatures captured by the heat equation model in the case
of the $\mathbf{T}_{r j}$ [also see Section~\ref{sec:SingleSite}(iii)].
Neither notion is unassailable: while the
errors in (\ref{eq:multipledatamodel}) are due to the measurement
process, they also respond to approximation errors associated
with the use of the reduced temperature definition. Similarly, the
errors in (\ref{eq:reduced_multipleprior}) are attributable to model
approximations associated with the specifics of our use of the heat
equation. However, we have virtually no prior information
regarding the structure of the unmodeled physical processes; if we did
know more, that knowledge could be used to improve the physical
models.

Ignorance is not a valid defense for independence
assumptions. Rather, those assumptions lead to obvious simplifications
in the analysis and avoid difficulties associated with potential
nonidentifiability issues. However, some sensitivity checks are
desirable. In our setting the major concern is that incorrect
assumptions of
independence may lead to underestimation of uncertainties in the
final results.

To assess independence assumptions, we examined model
``residuals.'' Some indication of structure may be developed by
inspecting estimated errors defined by
%
\begin{equation}
\label{eq:measerrorres1}
\hat{\mathbf{e}}_j = \mathbf{Y}_j -
[E(\mathbf{T}_{r j}) + T_{0 j} \mathbf{1}_{N_j} + E(q_{0 j})
\mathbf{R}_j],
 \qquad j=1,\ldots ,9,
\end{equation}
using estimated posterior expectations as indicated.
A more appropriate approach is to compute
%
\begin{equation}
\label{eq:measerrorres2}
\mathbf{e}_j^m = \mathbf{Y}_j -
[\mathbf{T}_{r j}^m + T_{0 j} \mathbf{1}_{N_j} + q_{0 j}^m
\mathbf{R}_j],  \qquad j=1,\ldots ,9,
\end{equation}
where the superscripts $m$ index MCMC iterations, though this option
leads to an ensemble of residuals.

\begin{figure}

\includegraphics{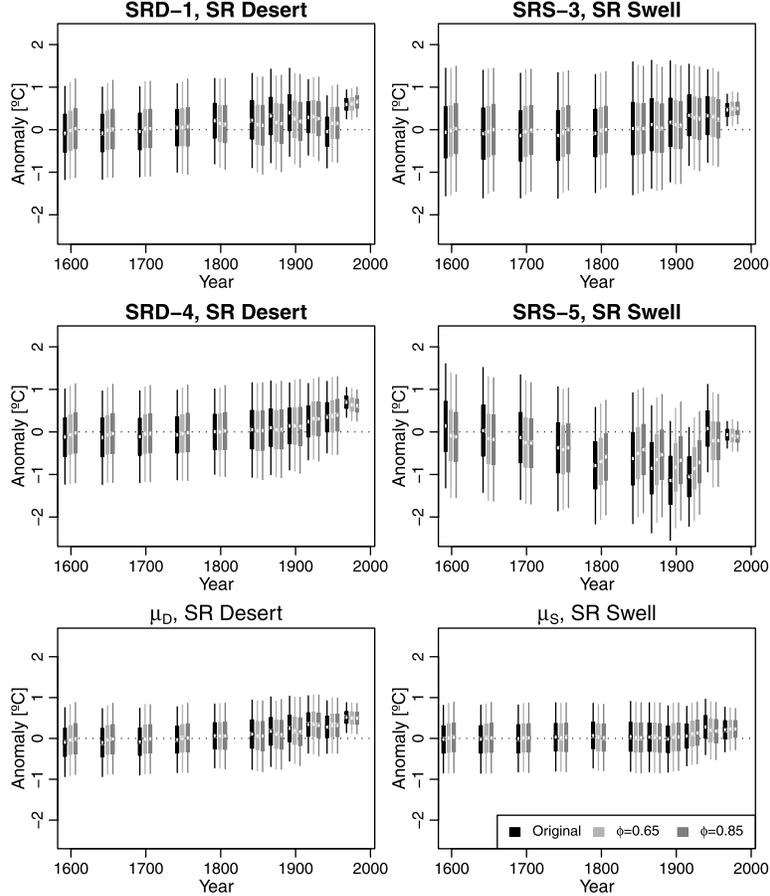}

\caption{Comparison of posterior distributions of
GST histories for
4 boreholes and the mean histories $\bmu_D$ and $\bmu_S$
using model error covariance matrices $\sigma_{j}^2 \mathbf{C}(\phi)$ with
different values of $\phi$.
The white squares show the estimated posterior means and the vertical bars
show the symmetric 50\% (thick) and 90\% (thin) posterior credible
intervals.
At each time point the first (darkest) bar
shows the original results ($\phi=0$; same as in Figure \protect\ref{fig:GTH})
and the next two bars show the results for
$\phi=0.65$ and $\phi=0.85$.}\label{fig:GTHspat}
\end{figure}

For each of the nine boreholes, we fitted time-series style ARMA models
to the $\hat{\mathbf{e}}_j$. Though we noted some differences,
AR(1) and ARMA(1,1) models provided reasonable fits. Since the forms of the
covariance matrices of the AR(1) and ARMA(1,1) are quite similar, we
focused on AR(1) models. We also inspected realizations of residuals
as defined in (\ref{eq:measerrorres2}). These
residuals were similar to those based on (\ref{eq:measerrorres1}),
though they exhibited less structure, suggesting that basing our
sensitivity checks on (\ref{eq:measerrorres1}) is conservative.
We replaced the covariances
in (\ref{eq:reduced_multipleprior}) by the covariance matrices
$\sigma_{j}^2 \mathbf{C}(\phi)$, where $\mathbf{C}(\phi)$ is a
correlation matrix with ones on the diagonal and off diagonal
elements $\phi^k$ for every pair of depths $k$ units apart (here one
unit corresponds to 5 meters). We treated $\phi$ as a known quantity
and reran the MCMC analyses for two choices of $\phi$: $0.65$
and $0.85$. The final
posterior results were not very different from those using the
independence assumption. In Figure \ref{fig:GTHspat} we display
comparisons of
posterior distributions of GST histories for two boreholes for each of
the Desert and Swell regions as well as the corresponding region mean
processes. The boreholes were selected to indicate the range of
differences, that is, one borehole showing the least differences and
another one
showing the most differences. We did
observe nonnegligible increases in the spreads of the posteriors for
the heat flow parameters $q_{0j}$, though there were virtually no changes
in their means.

We repeated this process by replacing the model error
covariances in (\ref{eq:multipledatamodel}) by matrices
$\sigma_{Y_j}^2 \mathbf{C}(\phi)$ for the same two values of
$\phi$. The same behaviors were observed; indeed, the differences
were smaller that those above. Hence, though some structure in the
errors are
unmodeled in this article, the effects of this appear to be very
minor in the posterior inferences regarding GST histories.\looseness=-1

\section{Discussion} \label{sec:Discussion}\label{sec5}
We developed Bayesian hierarchical models featuring two key aspects:
(1) the use of a physics based
model having uncertain parameters and subject to model error, and (2) the
illustration of borrowing strength\vadjust{\goodbreak} based on priors on selected parameters
to combine data sets. In the first development we relied on a common
framework to define reduced temperatures~$\mathbf{T}_r$ to justify use of
the heat equation as the main physical model. However, we added the
treatment of background surface heat flows $q_0$ as unknown parameters
and the inclusion of model errors implying a random heat equation
model. The hierarchical modeling approach also allows us to separate
explicitly measurement errors from model errors. These steps support
the claim that our models account for important uncertainties.

In our prior distributions we modeled the surface history processes and
background heat flows as arising from region-specific
(SR Desert or SR Swell) models, which in turn have parameters generated from
a San Rafael basin-wide prior model. These formulations led to
posterior results that display very notable and appropriate behaviors.
As discussed in Section~\ref{sec4.2}, the inferences for model parameters and
histories in the SR Desert region indicated borrowing strength
in concert with the similarities of behaviors
at individual boreholes. By comparison, we noted very little borrowing
of strength in the SR Swell region where the individual results were
comparatively dissimilar. We find this result quite
satisfactory and illustrative, but should note that the priors used
were extremely simple. They were chosen because we had little prior
information and too little data
(i.e., four and five boreholes in the two regions)
on which to update more intense priors. When feasible,
we recommend consideration of more intense spatial process priors for
parameters. See \citet{Cressie1993} and \citet{BanerjeeCarlinGelfand} for
discussion of
sophisticated spatial models.

As is appropriate in most Bayesian analyses, we devoted substantial
attention to sensitivity analyses. The results are discussed in
Section~\ref{sec4} and not reviewed here. However, note that we did not present
analyses regarding the specification of the thermal conductivities
($k$) used in defining the vectors of thermal resistance vectors
$\mathbf{R}$. Very cursory inspections suggest to us that the approach may
be very sensitive to these quantities. We will pursue this aspect in an
alternative modeling approach to be reported on elsewhere.

We note that the results lead to the
suggestion that borehole data are useful in inferring surface
temperatures for times from the recent past to about 200 years in the
past. For times deeper in the past, the borehole data appear to be
comparatively less informative. We base this suggestion on the
behavior of the posterior distributions of the site-wise histories
as well as the SR Desert and Swell mean histories. These distributions
seem to
asymptote to region-specific distributions. For all times before 1800
and all five boreholes in the SR Desert, the posterior standard
deviations of historical temperatures vary between $0.61$ and $0.68$
(only 5 of the 35 values are less than $0.65$); these values are 2 or
3 times larger than the standard deviations for temperatures in the
Desert in 1980. In the SR Swell, all 28 of the corresponding
standard deviations are between $0.87$ and $0.95$, and are roughly 4
times the standard deviations for temperatures in the Swell in
1980. Regarding this issue, \citeauthor{NorthEtal2006} [(\citeyear{NorthEtal2006}), page~80] write the following:

\begin{quote}

The time resolution and length of borehole-based surface temperature
reconstructions are severely limited by the physics of the heat
transfer process\ldots . A surface temperature signal is irrecoverably
smeared as it is transferred to depth. The time resolution of the
reconstruction thus decreases backward in time.
For rock and permafrost boreholes, this resolution is a few decades at
the start of the 20th century and a few centuries at 1500.

\end{quote}

 \noindent
For related discussion see \citet{BeltramiMareschal1995} and
\citet{HopcroftGallagherPain2007}.
Our addition to these claims is that the posterior standard deviations
based on models that include model error and other uncertainties are
roughly constant
by necessity for times beyond 200 years in the past. Of course, this
is based on limited data from a limited region and need not apply in
greater generality.

\begin{figure}[b]

\includegraphics{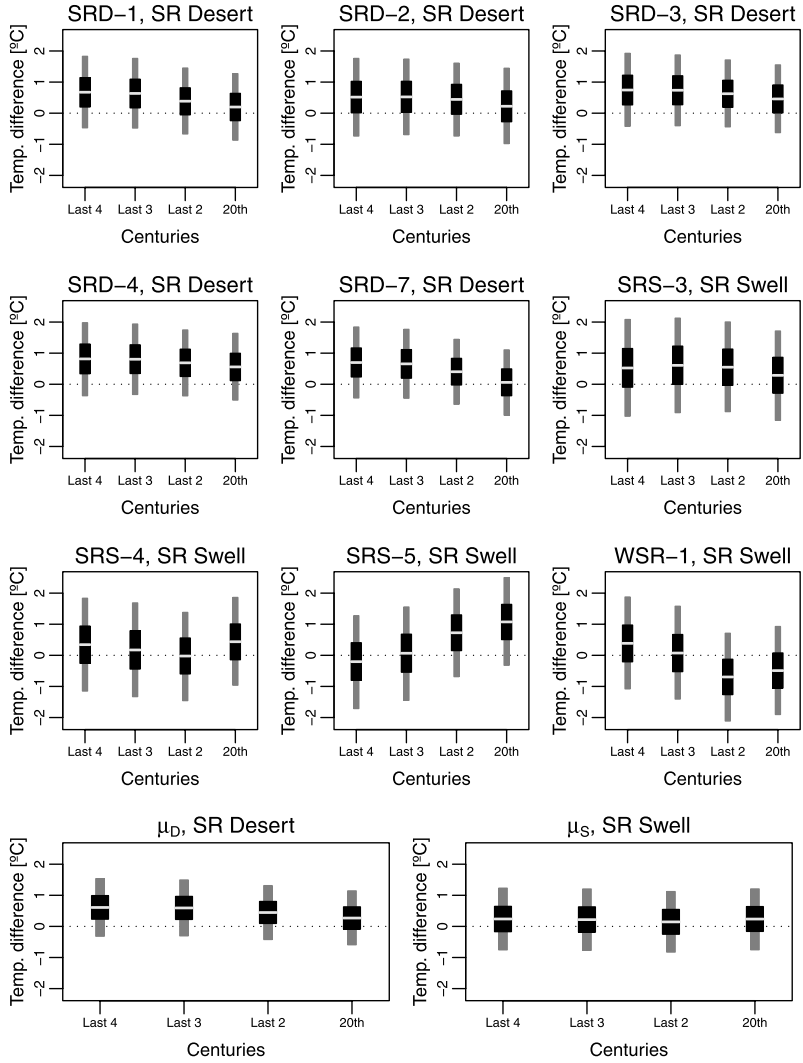}

\caption{Posterior means and credible sets of
surface temperature changes over the four
periods 1600, 1700, 1800 and 1900 to the latest year in
our data sets (1980) for each of the nine boreholes and for the SR Desert
and SR Swell means. The vertical bars
show the symmetric 50\% (thicker) and 90\% (thinner) posterior credible
intervals and the grey horizontal lines show the posterior means.}\label{fig:TempDiff}
\end{figure}

To characterize our results in regard to climate change, we provide
inferences on the changes in surface temperatures over the four
periods 1600, 1700, 1800 and 1900 to the latest year in
our data sets (1980) for each of the nine boreholes and for the SR Desert
and SR Swell means (see Figure~\ref{fig:TempDiff}). We note that the
point estimates of
these changes are typically positive (i.e., increased temperature) for
the individual boreholes and are all positive for the basin-wide
means. However, all $90\%$ credible intervals cover~$0^{\circ}$C,
though many of the $50\%$ credible intervals lie above $0^{\circ}$C.
Based on the common trend in these
results, we believe that the suggestion of warming is supported.
While the strength of this support is not strong, we
note that our sample sizes are very small.
Further, since our data ends in 1980, we cannot find
the more recent warming reflected in other data. Finally, we
caution that traditional quantification associated with so-called
\textit{statistical significance} (i.e., $90\%$ or $95\%$ intervals not
covering 0) are of little relevance in regard to decision making in
the context of climate change.

\appendix
\section{\texorpdfstring{List of parameters}{Appendix A: List of parameters}} \label{app:List}
\textit{Unknown parameters}:

Borehole specific parameters ($j=1, \ldots , 9$):
\begin{itemize}[$\sigma^2_{Yj}$:]
\item[$\mathbf{T}_{hj}$:]   Temperature histories, vectors of dimension
$K$.
\item[$\mathbf{T}_{rj}$:]   True reduced temperatures, vectors of dimension
$N_j$.
\item[$q_{0j}$:]   Heat flows, scalars.
\item[$\sigma^2_{Yj}$:]   Measurement error variances.
\item[$\sigma^2_j$:]   Model error variances.
\end{itemize}

Region specific parameters:
\begin{itemize}[$\bmu_D$, $\bmu_S$:]
\item[$\bmu_D$, $\bmu_S$:]   Mean temperature histories for SR Desert (D) and  SR Swell (S), vectors of dimension
$K$.
\item[$\gamma^2_D$, $\gamma^2_S$:]   Variances of temperature histories for SR
Desert (D)   and SR Swell (S).
\item[$\nu_D$, $\nu_S$:]  Mean heat flow for SR Desert (D) and Swell (S),
scalars.
\item[$\tau^2_D$, $\tau^2_S$:]   Variances of heat flow for SR Desert (D) and
Swell (S).
\end{itemize}

\textit{Hyperparameters--constants}:
\begin{itemize}[$\sigma^2_0$, $\sigma^2_D$, $\sigma^2_S$:]
\item[$\bmu_0$:]   Prior mean of $\bmu_D$ and $\bmu_S$.
\item[$\sigma^2_0$, $\sigma^2_D$, $\sigma^2_S$:]   Define the prior covariance
structure of $\bmu_D$ and $\bmu_S$.
\item[$\nu_0$:]   Prior mean of $\nu_D$ and $\nu_S$.
\item[$\eta^2_0$, $\eta^2_D$, $\eta^2_S$:]   Define the prior covariance
structure of $\nu_D$ and $\nu_S$.
\item[$a_Y$, $b_Y$:]   Define the Inverse Gamma prior for
$\sigma^2_{Yj}$.
\item[$a$, $b$:]   Define the Inverse Gamma prior for $\sigma^2_{j}$.
\item[$a_\gamma$, $b_\gamma$:]   Define the Inverse Gamma prior for $\gamma
^2_{D}$ and $\gamma^2_{S}$.
\item[$a_\tau$, $b_\tau$:]   Define the Inverse Gamma prior for $\tau^2_D$ and
$\tau^2_S$.
\end{itemize}

\section{\texorpdfstring{Gibbs sampler}{Appendix B: Gibbs sampler}} \label{app:Gibbs}
A Gibbs sampler is a method that obtains approximate samples from the
posterior distribution. It avoids
the big dimensionality of the model by simulating only parts of the
parameters at a time,
using the so-called full conditional distributions. The notation $
[X | \cdot ] $ reads ``the full conditional
distribution of $X$ given all other parameters and the data.'' Also,
$\| \mathbf{x} \| = \mathbf{x}^\prime\mathbf{x} $, for a vector
$\mathbf{x}$.

\subsection*{GST history vectors $\mathbf{T}_{hj}$ and reduced temperature
vectors $\mathbf{T}_{rj}$}

To make the Gibbs sampler more efficient, we sample the joint full
conditional distribution
$\mathbf{T}_{hj}, \mathbf{T}_{rj} | \cdot$ for each borehole $j$.
Note that the joint (prior) distribution of $\mathbf{T}_{hj}$ and
$\mathbf{T}_{rj}$ for boreholes in the SR desert
($j=1,\ldots ,5$) is
%
\begin{equation} \label{eq:TfullPrior}
 \pmatrix{\displaystyle
\mathbf{T}_{rj} \cr\displaystyle
\mathbf{T}_{hj}
}
 \sim N  \left(
\pmatrix{\displaystyle
A_j \bmu_D \cr\displaystyle
\bmu_D
}
 ,
 \pmatrix{\displaystyle
\widetilde{\Sigma}_j & \gamma^2_D A_j \cr\displaystyle
\gamma^2_D A_j^\prime& \gamma^2_D I_K
}
\right  ),
\end{equation}
where
%
\begin{equation} \label{eq:SigmaTilde}
\widetilde{\Sigma}_j = \sigma_j^2 I_{N_j} + \gamma^2_D A_j A_j^\prime .
\end{equation}
The joint full conditional distribution of $\mathbf{T}_{rj}$ and
$\mathbf{T}_{hj}$ given $\mathbf{Y}_j$ and all
other parameters, which we denote by $\Theta$, can be written as follows:
%
\begin{eqnarray}
\label{eq:TjFull}
 [ \bT_{rj}, \bT_{hj} | \bY_j, \Theta ] & =&
\frac{ [ \bY_j |\bT_{rj} ,\Theta  ]
 [\bT_{rj} | \Theta  ]} { [\bY_j| \Theta  ]}
 [ \bT_{hj} | \bT_{rj} ,\Theta  ]
\nonumber
\\[-8pt]
\\[-8pt]
& =&  [ \bT_{rj} |\bY_j ,\Theta  ]
 [\bT_{hj} | \bT_{rj} , \Theta  ].
\nonumber
\end{eqnarray}
Therefore, to sample
$\mathbf{T}_{hj}, \mathbf{T}_{rj} | \cdot$,
we first sample the marginal distribution $ [ \bT_{rj} |\bY_j
,\Theta  ] $ and then the
conditional distribution $ [\bT_{hj} | \bT_{rj} , \Theta  ] $.
First note that
$ [ \bT_{rj} |\bY_j ,\Theta ]$
is proportional to $ [ \bY_j | \bT_{rj} , \Theta ]  [ \bT_{rj} |
\Theta ]$ where $[ \bT_{rj} |
\Theta ]$ is the marginal
distribution of the joint prior (\ref{eq:TfullPrior}). Therefore, for
boreholes in the SR Desert
($j=1, \ldots , 5$), we have $ [ \bT_{rj} |\bY_j ,\Theta  ] =
N(Dd, D)$, where
%
\begin{equation}
D =  \biggl(\frac{1}{\sigma_{Yj}^2} I_{N_j} + \widetilde{\Sigma
}_j^{-1} \biggr)^{-1}
\end{equation}
and
%
\begin{equation}
\label{eq:TrPost}
d =  \bigl(\bY_j - (T_{0j}\mathbf{1}_{N_j} + q_j\mathbf{R}_j) \bigr)/\sigma
_{Yj}^2 + \widetilde{\Sigma}_j^{-1}A_j \bmu_D .
\end{equation}
Second, note that $ [\bT_{hj} | \bT_{rj} , \Theta  ]$ is the
conditional prior distribution
%
\begin{equation}
N  \bigl( \bmu_D + \gamma^2_D A_j^\prime\widetilde{\Sigma}_j^{-1}
(\mathbf{T}_{rj} - A_j \bmu_D),
\gamma^2_D I_K - \gamma^4_D A_j^\prime\widetilde{\Sigma}_j^{-1} A_j
 \bigr).
\label{eq:ThPost}
\end{equation}
For boreholes in the SR Swell ($j = 6, \ldots , 9$), we replace $\bmu_D$
in (\ref{eq:TrPost}) and (\ref{eq:ThPost})
with $\bmu_S$ and $\gamma^2_D$ in (\ref{eq:ThPost}) and (\ref
{eq:SigmaTilde}) with $\gamma^2_S$.

\subsection*{Measurement and model error variances, $\sigma^2_{Yj}$ and
$\sigma^2_j$}

%
\begin{eqnarray}\label{eq:sigmaY}
\sigma^2_{Yj} | \cdot & \sim&
IG  \biggl( \frac{N_j}{2} + a_Y,
b_Y + \frac{1}{2}
 \| \mathbf{Y}_j-(\mathbf{T}_{rj}+T_{0j}\mathbf{1}_{N_j} +
q_j\mathbf{R}_j)  \|
 \biggr),  \\
\label{eq:sigmar}
\sigma_j^2 | \cdot& \sim&
IG  \biggl( \frac{N_j}{2} + a,
b + \frac{1}{2} \| \mathbf{T}_{rj}-A_j\mathbf{T}_{hj}
\|  \biggr), \qquad j=1, \ldots , 9.
\end{eqnarray}

\subsection*{Mean and variances of the GST history, $\bmu_D$, $\bmu_S$,
$\gamma^2_D$ and $\gamma^2_D$}
We sample the joint full conditional distribution of $\bmu_D$ and $\bmu
_S$. We have
$\bmu_D, \bmu_S | \cdot\sim N(Dd, D)$ with
%
\begin{equation}
D = \frac{1}{w}
\pmatrix{\displaystyle \biggl( \frac{4}{\gamma^2_S} + \frac{\sigma^2_D + \sigma
^2_0}{v}  \biggr) I_K& \displaystyle\frac{\sigma^2_0}{v}I_K \cr\displaystyle
\frac{\sigma^2_0}{v} I_K & \displaystyle \biggl(\frac{5}{\gamma^2_D} + \frac{\sigma
^2_S + \sigma^2_0}{v}  \biggr) I_K
}
\end{equation}
and
%
\begin{equation}
d =
\pmatrix{\displaystyle \frac{1}{\gamma^2_D}\sum_{j=1}^5 \mathbf{T}_{hj} \cr\displaystyle
\frac{1}{\gamma^2_S}\sum_{j=6}^9 \mathbf{T}_{hj}
}
+ \frac{1}{v}
\pmatrix{\displaystyle \sigma^2_S \bmu_0 \cr\displaystyle  \sigma^2_D \bmu_0
}
,
\end{equation}
where $w =  ( \frac{5}{\gamma^2_D} + \frac{\sigma^2_S + \sigma
^2_0}{v}  )
 ( \frac{4}{\gamma^2_S} + \frac{\sigma^2_D + \sigma^2_0}{v}
) - \frac{\sigma^4_0}{v^2} $
and $v= (\sigma^2_D + \sigma^2_0)(\sigma^2_S + \sigma^2_0) - \sigma^4_0$.

The variances of the GST histories are sampled separately for each region,
%
\begin{equation} \label{eq:gammaDPost}
\gamma_D^2 | \cdot \sim IG  \Biggl( 5K/2 + a_\gamma, b_\gamma+
\frac{1}{2} \sum_{j=1}^5  \|\mathbf{T}_{hj} -\bmu_D
\|  \Biggr)
\end{equation}
and
%
\begin{equation}
\label{eq:gammaSPost}
\gamma_S^2 | \cdot \sim IG  \Biggl( 4K/2 + a_\gamma, b_\gamma+
\frac{1}{2} \sum_{j=6}^9  \|\mathbf{T}_{hj} -\bmu_S
\|  \Biggr).
\end{equation}

\subsection*{Heat flow parameters $q_{0j}$, $\nu_D$, $\nu_S$, $\tau^2_D$,
$\tau^2_S$}

For boreholes in the SR Desert ($j=1, \ldots , 5$), we have
%
\begin{equation}
\label{eq:qPost}
q_j | \cdot  \sim
N \biggl(
\frac{\tau_D^2 \mathbf{R}_j^\prime (\mathbf{Y}_j- \mathbf{T}_{rj}
- T_{0j}\mathbf{1}_{N_j} )
+ \sigma_{Yj}^2 \nu_D }
{ \tau_D^2 \mathbf{R}_j^\prime\mathbf{R}_j + \sigma_{Yj}^2 } ,
\frac{\tau_D^2 \sigma_{Yj}^2}{ \tau_D^2 \mathbf{R}_j^\prime\mathbf
{R}_j + \sigma_{Yj}^2 }  \biggr) .
\end{equation}
For boreholes in the SR Swell ($j= 6, \ldots , 9$), we replace $\bmu_D$
and $\tau^2_D$ in (\ref{eq:qPost}) by
$\bmu_S$ and $\tau^2_S$.

The mean heat flow parameters $\nu_D$ and $\nu_S$ are sampled from a
joint distribution
$\nu_D, \nu_S | \cdot\sim N(Dd, D)$ with
%
\begin{equation}
D = \frac{1}{w}
\pmatrix{\displaystyle\frac{4}{\tau^2_S} + \frac{\eta^2 + \eta^2_0}{v} &\displaystyle \frac
{\eta^2_0}{v} \cr\displaystyle
\frac{\eta^2_0}{v} &\displaystyle \frac{5}{\tau^2_D} + \frac{\eta^2 + \eta^2_0}{v}
}
\end{equation}
and
%
\begin{equation}
d =
\pmatrix{\displaystyle \frac{1}{\tau^2_D} \sum_{j=1}^5 q_j \cr\displaystyle  \frac{1}{\tau
^2_S} \sum_{j=6}^9 q_j
}
+ \frac{1}{v}
\pmatrix{\displaystyle \eta^2 \nu_0 \cr\displaystyle  \eta^2 \nu_0
}
,
\end{equation}
where $w =  (\frac{5}{\tau^2_D} + \frac{\eta^2 + \eta^2_0}{v}  )
 (\frac{4}{\tau^2_S} + \frac{\eta^2 + \eta^2_0}{v}  ) -
\frac{\eta^4_0}{v^2}$ and $v = (\eta^2 + \eta^2_0)^2 - \eta^4_0$.

Finally, the variances of the heat flow are sampled separately for each region,
%
\begin{equation} \label{eq:tauDPost}
\tau_D^2 | \cdot \sim
IG \Biggl ( 5/2 + a_\tau, b_\tau+ \frac{1}{2} \sum_{j=1}^5 (q_j-\nu
_D)^2  \Biggr)
\end{equation}
and
%
\begin{equation}
\label{eq:tauSPost}
\tau_S^2 | \cdot  \sim IG  \Biggl( 4/2 + a_\tau, b_\tau+ \frac
{1}{2} \sum_{j=6}^9 (q_j-\nu_S)^2  \Biggr) .
\end{equation}
%

\section*{Acknowledgments}
We thank the Editor Michael Stein, the Associate Editor and three
anonymous referees
whose comments greatly improved the paper.
We thank Dr. Robert Harris for providing us with the data
and for helpful discussions. Jenn\'{y} Brynjarsd\'{o}ttir is grateful
to Dr. Doug Wolfe for
support during the development of this article.


%

\printaddresses

\end{document}